# Neural correlates of cognitive ability and visuo-motor speed: validation of IDoCT on UK Biobank Data

Valentina Giunchiglia[1], Sharon Curtis[2], Stephen Smith[3], Naomi Allen[2], Adam Hampshire[1]

[1] Department of Brain Sciences, Imperial College London, London, United Kingdom,
[2] Nuffield Department of Population Health, Medical Sciences Division, University of Oxford
[3] Wellcome Centre for Integrative Neuroimaging (WIN FMRIB), Oxford University, Oxford, United Kingdom

**Abstract**

Automated online and App-based cognitive assessment tasks are becoming increasingly popular in large-scale cohorts and biobanks due to advantages in affordability, scalability and repeatability. However, the summary scores that such tasks generate typically conflate the cognitive processes that are the intended focus of assessment with basic visuomotor speeds, testing device latencies and speed-accuracy tradeoffs. This lack of precision presents a fundamental limitation when studying brain-behaviour associations. Previously, we developed a novel modelling approach that leverages continuous performance recordings from large-cohort studies to achieve an iterative decomposition of cognitive tasks (IDoCT), which outputs data-driven estimates of cognitive abilities, and device and visuomotor latencies, whilst recalibrating trial-difficulty scales. Here, we further validate the IDoCT approach with UK BioBank imaging data. First, we examine whether IDoCT can improve ability distributions and trial-difficulty scales from an adaptive picture-vocabulary task (PVT). Then, we confirm that the resultant visuomotor and cognitive estimates associate more robustly with age and education than the original PVT scores. Finally, we conduct a multimodal brain-wide association study with free-text analysis to test whether the brain regions that predict the IDoCT estimates have the expected differential relationships with visuomotor vs. language and memory labels within the broader imaging literature. Our results support the view that the rich performance timecourses recorded during computerised cognitive assessments can be leveraged with modelling frameworks like IDoCT to provide estimates of human cognitive abilities that have superior distributions, re-test reliabilities and brain-wide associations.

## 1. INTRODUCTION

Automated and app-based assessment technologies provide a scalable, cost-effective, and reliable way to measure different aspects of cognitive abilities (1) and to monitor cognitive changes in clinical populations (2–4). Consequently, this technology is becoming popular in large-scale citizen science projects (5,6), cohorts (7) and registers (8). Building on the resultant big data, a major research drive has been to map associations between the summary scores that these computerised cognitive tasks output, and features of brain structure and function from large-scale imaging cohort studies (9,10). However, it is common practice to summarise a participants' performance by estimating or contrasting average accuracy and reaction times (RT) across task conditions (11). The resultant scores relate not only to individual differences in abilities to process the specific cognitive demands that are the intended target of the task (12,13), but also to other confounding factors such as visuo-motor processing speeds and the latency of the devices that people are assessed with. This lack of cognitive precision in summary score estimates is a non-trivial limitation for both the strength and the specificity of the associations that can be achieved.

A commonly overlooked advantage of computerised cognitive tasks is that, unlike classic pen-and paper assessment scales, they record every stimulus and response in a detailed performance timecourse. These performance timecourses can be modelled in more sophisticated ways to obtain ability estimates that have superior reliability and process specificity than simple contrasts or averaging. Previously, we reported development of one such modelling framework, IDoCT (Iterative Decomposition of Cognitive Tasks – see Box 1), which we designed to disentangle individuals' cognitive abilities from other confounding factors in a manner that is robust, computationally inexpensive and sufficiently flexible to be adapted for practically any task that manipulates cognitive difficulty across trials (14).

More specifically, IDoCT leverages trial-by-trial performance data from large cohorts to (a) estimate individuals' abilities to specifically cope with higher cognitive difficulty across trials (AS) and estimate their visuo-motor delay times (DT) while accounting for individual speed-accuracy tradeoffs. Notably, the approach concurrently recalculates the relative difficulty assigned to trials across task conditions using a fixed-point iterative process that handles the circularity of simultaneously defining individual performance from trial difficulty and trial difficulty from individual performance (14). This is achieved in a manner that accounts for any bias towards sampling more difficult conditions in more able individuals, as is the case in some adaptive designs, resulting in a robust data-driven recalibration of trial-difficulty scales.

We initially validated IDoCT by applying it to data from >400,000 participants who undertook 12 cognitive tasks during the Great British Intelligence Test (6). The results showed a successful decomposition of cognitive vs. device and visuomotor latencies, as gauged by superior sociodemographic associations and AS estimates that were not dwarfed by an inflated global intelligence factor (14). This combination of sensitivity and decorrelation holds promise for achieving stronger and more process-specific functional-anatomical mappings in behavioural-imaging association studies.

Here, we further validate IDoCT in the context of an adaptive Picture Vocabulary Task (PVT) (15,16) that was designed to measure 'crystallised' comprehension and reading decoding abilities (8,17), and that was deployed with 34,927 participants as part of the UK BioBank (UKB) imaging extension (18). Extracting summary measures for this PVT presents a notable challenge because it applied an adaptive staircase sampling algorithm to efficiently measure each participant's ability level, but the word-picture difficulty scale that the sampling algorithm traversed proved to be sub-optimally calibrated for the assessed population. This resulted in aberrant sampling trajectories with accuracy ceiling effects and an unexpected bimodal distribution of population performance scores that is not ideal for analysis of associations (18). In theory, IDoCT should be able to resolve this ill-posed problem by recalibrating the trial-difficulty scale whilst using both speed and accuracy to produce more precise estimates of participant performance, with applications in functional-brain mapping studies.

To test this theory, we first confirm the expected improvements in distributions and test-retest reliability of PVT performance estimates for IDoCT relative to the original summary scores. Next, we evaluate whether the IDoCT PVT performance estimates associate more robustly, and in an interpretable manner, with participant age and education. Then, we use a simple linear machine learning pipeline to test the hypothesis that the IDoCT PVT estimates can be more reliably predicted from four distinct feature sets of the UKB structural imaging database. Finally, we conduct free text mining across the imaging literature to determine whether the brain regions that predict IDoCT PVT estimates of visuomotor and cognitive abilities are differentially associated in the neuroscience literature with visual and motor functions vs. language and memory functions respectively.

**Box 1. IDoCT: Iterative Decomposition of Cognitive Tasks**

1. **Data driven assessment of trial's difficulty D(t)**

The performance P(i,t) of participant i in trial t is calculated based on the RT of participant *i* in trial *t* and the difficulty D(t) of trial *t*, if the answer is correct. In case of incorrect answers, the performance is equal to 0.

$$P(i,\ t) = \begin{cases} 0, & if\ wrong \\ \left(1 - \frac{RT(i,t)}{RT_{max}}\right) * D(t), & if\ right \end{cases}$$

D(t) is calculated according to the performance P(i,t) across all participants *i* in trial *t*

$$D(t) = \frac{1}{\sum_{i=1}^{N} T(t,\ i)} \sum_{i=1}^{N} \sum_{T(t,\ i)=1}^{T(t,i)} 1 - P(i,\ t)$$

A mutual recursive definition is generated. At the first iteration D(t) is set equal to 1 for all trials *t*, and then iteratively modified. The iterations are interrupted when the model converges into an invariant measure of trial's difficulty D(t).

2. **Data driven assessment of answer time AT(i,t)**

RT is characterized by two components: answer time AT, or cognitive time required to provide an answer to the cognitive task, and DT, or visuo-motor latency time. The measure of performance can be updated as follows:

$$P(i,\ t) = \begin{cases} 0, & if\ wrong \\ \left(1 - \frac{AT(i,t) + DT(i,t)}{RT_{max}}\right) * D(t), & if\ right \end{cases}$$

The answer time AT(i,t) of participant *i* in trial *t* is calculated according to the ability A(i) of the participant, the RT(i,t) and the difficulty of the trial D(t).

$$ATN(i,t) = \big(1 - A(t)\big) * D(t)$$
$$AT(i,t) = ATN(i,t) * (RT(i,t) - DT_{MAX}(i))$$

Where

$$DT_{max}(i) = RT(i,t)\ for\ the\ t\ such\ that\ RT(i,t)\ is\ the\ smallest$$

The ability A(i) is measured based on the cumulative performance of participant *i* across all trials *t.*

$$B(i,t) = B(i,t-1) + P(i,t)$$
$$BN(i,N) = \frac{\sum_{t=1}^{N} B(i,t)}{N}$$
$$A(i) = BN(i,Q)$$

Where B(i,t) corresponds to the cumulative performance of participant *i* up to trial *t*, and BN(i,N) to the overall average performance across all N trials completed.
The delay time DT(i,t) is calculated as the difference between the reaction time RT(i,t) and the answer time AT(i,t).

$$DT(i,t) = RT(i,t) - AT(i,t)$$

A mutual recursive definition between A(i) and P(i,t) is generated. The first iteration assumes that A(i) is maximum for all participants i and is, therefore, always equivalent to 1, the measure of AT is initialized as ATmin(i) (which is measured as the difference between RT and $DT_{max}$). Instead, DT(i, 0) is initialized as DTmax(i). DTmax(i) is set equal to RTmin(i). The iterations are interrupted when the model converges.

3. **Measure of specific ability AS(i) and delay time DT(i)**

Specific ability AS(i) is calculated according to the cumulative specific performance PA(i,t) of participant *i,* which is measured using the RT(i,t) corrected for the DT(i,t).

$$PA(i,t) = \begin{cases} 0, & if\ wrong\ answer \\ \left(1 - \frac{AT(i,t)}{AT_{max}}\right) D(t), & if\ right\ answer \end{cases}$$

The delay time DT(i) is measured as the average delay time D(i,t) across all N trials *t*.

$$DT(i) = \sum_{t=1}^{N} \frac{DT(i,t)}{N}$$

4. **Measure of scaled trial difficulty DS(i)**

The trial difficulty D(t) is scaled according to the specific ability AS(i) of all the N participants *i* that completed the trial *t.*

$$DS(t) = \frac{\sum_{i=1}^{N} AS(i) * D(t)}{N}$$

## 2. MATERIALS AND METHODS

Our full analysis pipeline consisted of 9 steps, summarized in Figure 1, namely 1) data curation, 2) IDoCT modelling, 3) evaluation of distributions, 4) estimation of trials' difficulty trajectories, 5) assessment of test-retest reliability, 6) association with age and education, 7) imaging associations, 8) automated literature extraction and preprocessing, and 9) free text analysis.

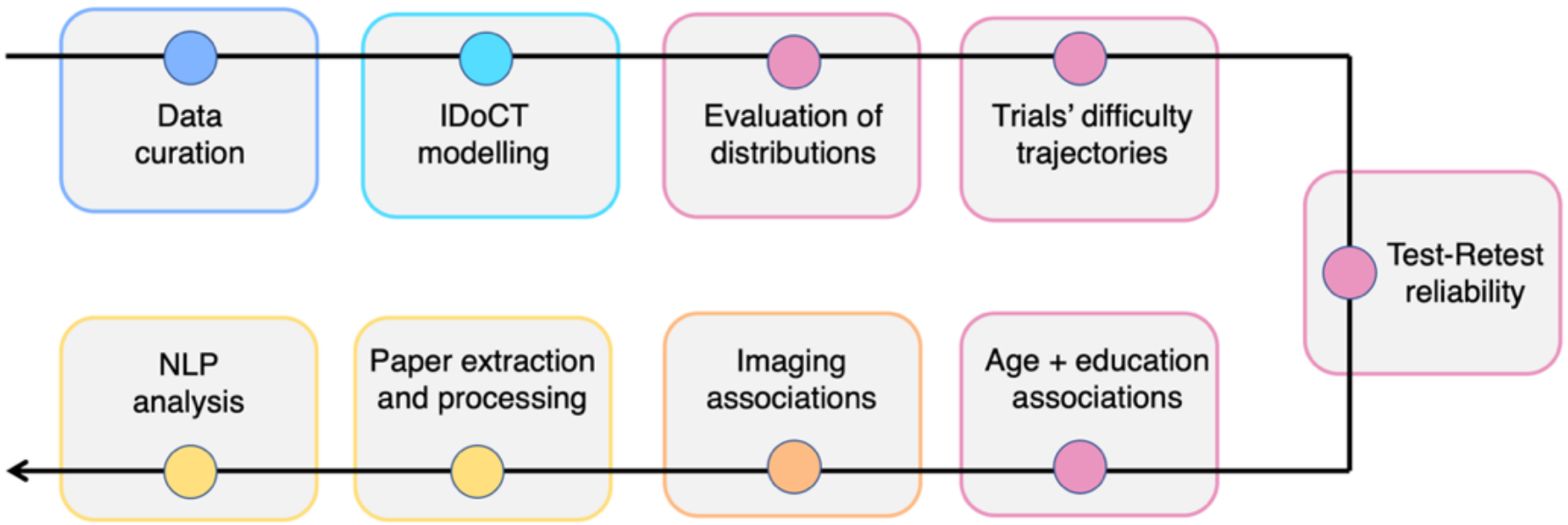

**Figure 1 Summary of analysis.** The full analysis consisted of 9 steps: 1) data curation, 2) IDoCT modelling, 3) evaluation of distributions, 4) estimation of trials' difficulty trajectories, 5) assessment of test-retest reliability, 6) association with age and education, 7) imaging associations, 8) automated literature extraction and preprocessing, and 9) free text analysis.

### 2.1 Study design and participants

The data analysed in this study were collected as part of UKB, a population-based prospective study that recruited >500,000 participants, aged between 40 and 69, in the 2010-2016 timeframe. The aim of UKB is to understand the genetic and non-genetic factors that contribute to different diseases that affect mainly the middle and older aged population. As part of the study, a subset of individuals undertook imaging, genetic, as well as health and demographics measures longitudinally, across 22 assessment centers throughout the United Kingdom (18). All participants provided informed, written, consent. Detailed information on the study design is available in the original paper (10) or online https://www.ukbiobank.ac.uk/. In the current study, demographics (i.e. age and education), imaging (i.e. fractional anisotropy from diffusion weighted images, and volume, intensity and thickness measures from structural MRI), and behavioural data (i.e. performance in PVT) were used.

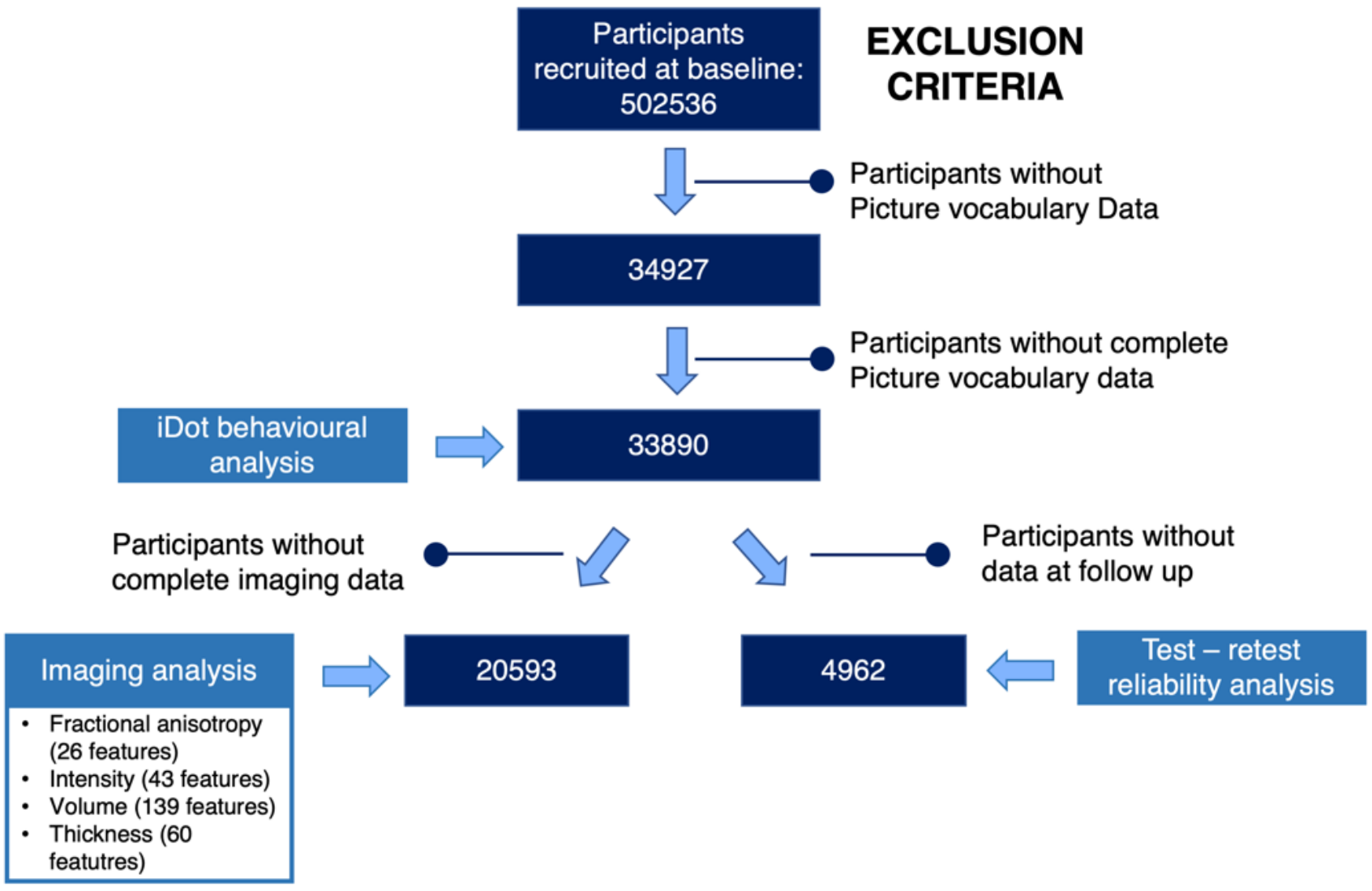

**Figure 2. Sample and data flowchart.** From the 502536 participants in the UK Biobank, 33890 participants had complete data for the Picture Vocabulary data, among which 20593 had data for all imaging modalities of interest, which consisted of 26 fractional anisotropy features from diffusion weighted images, 43 intensity measures, 139 volume and 60 thickness measures from structural MRI. 4962 participants had picture vocabulary data at follow up.

In total, 34927 participants completed the PVT during at least one timepoint. 33890 had complete picture vocabulary task data at baseline, of whom 20593 had data for all imaging measures of interest, and 4962 had cognitive data at follow up. Figure 2 shows the exact numbers of participants used for each step of the analysis and the exclusion criteria.

## 2.2 Picture Vocabulary task (PVT)

The PVT used for the UK Biobank was adapted from the NIH Toolbox Picture Vocabulary Test (15,16,19). This task was designed to assess a person's semantic/language functions by measuring their ability to match words to pictures. At each trial, participants were presented with a written word along with a set of 4 images, and they were required to select the picture that matched the word. Every participant was started on the same word, which was at an easy vocabulary level. The difficulty of the trials was then adaptively changed, using an Item Reponse Theory (IRT) model, in response to the participant's performance, according to a maximum likelihood estimate (MLE) of their vocabulary level, calculated from their answers so far. For example, if a participant selected the correct picture at trial n, then the word-picture combination presented at trial n+1 was at a higher level of difficulty, unless the supply of more difficult words had run out. Each participant's sequence of trials lasted for at least 20 words, and was terminated when either the accuracy of the estimate was close enough (using the standard error value), or a maximum of 30 trials had been reached.

The similarities and differences from the NIH toolbox test were as follows. The UK Biobank version used a touchscreen to present the words in written form only, unlike the NIH version administered by staff who also pronounced the words. The dataset of pictures, words, and their associated difficulty levels, was the same as the English-language version of the NIH PVT, but with modification for obvious difficulties: words differing between UK and US English were changed or removed, for example the "intersection" (US) was replaced with "crossroads" (UK), and the word "minute" was removed due to ambiguity of its written meaning, as the picture of a clock could have detracted from its intended meaning of a small size. This resulted in a dataset for 340 words. The same methods for estimating vocabulary levels and algorithmic adaptation for the next question were used in both tests. However the starting word and the maximum number of trials differed for the UK Biobank test.

## 2.3 Processing of UKB imaging data

The imaging data consisted of 26 measures of fractional anisotropy (FA) from diffusion weighted images (DWI), and 139 volume, 43 intensity and 60 cortical thickness measures from T1-weighted structural magnetic resonance images (MRI). All these imaging features were provided by UKB. Detailed information on the imaging processing is available in (20) and UKB feature labels are provided in the Supplementary Table A1. In brief, DICOM images were converted to NIFTI, fully anonymized using a defacing mask, and corrected for gradient distortion (GD). Specifically, in case of T1 images, the field of view (FOV) was cut down, the images were non-linearly registered to the standard MNI152 space (1mm resolution) and the brain was extracted. Finally, tissue segmentation was conducted to identify the different tissues and subcortical structures, and final volume measures were extracted after correcting for total brain size. In case of diffusion weighted images (dMRI), the first step of the processing consisted of the correction for head motion and eddy currents, followed by gradient distortion correction. Then, measures of fractional anisotropy (FA) and mean diffusivity were extracted.

## 2.4 IDoCT: extraction of measures of visuo-motor latency and cognitive ability

IDoCT is an iterative-based method (14) that takes as input trial-by-trial measures of reaction time (RT), and of accuracy for each participant, which can be binary (i.e. whether the participant replied correctly or not in each trial) or continuous (i.e. how correct was each individual response). In addition, it requires condition labels for each trial in the timecourse, which are defined based on the design of the task. Here, for example, the words presented at each trial are assumed to vary in difficulty; therefore, they are each assigned their own unique condition label. Then, through two separate iterative processes, the model returns measures of trial difficulties (D), of specific ability (AS) and of basic visuo-motor response speed (DT) for each participant. In a further step, it calculates scaled measures of trial difficulties (DS), that account for cases in which the trial assignment across the participants of differing abilities was biased towards different difficulty levels. This occurs, for example, when the most difficult trials are presented exclusively or predominantly to the most able participants, which should be the case here if the PVT sampling algorithm has sampled across an ordered difficulty scale. DS is obtained by scaling D based on the ability of the participants that completed the specific trials. Details on the implementation of IDoCT are available in the original publication (14) and are summarised in Box 1. In case of PVT, the accuracy measures are binary, and each trial is defined according to the word that is presented to the participants. Therefore, a given trial could occur just once per participant per session, with different participants completing different combinations of trials.

### 2.5 Test-Retest reliability

IDoCT estimates were computed on the follow up data to assess the test-retest reliability of the model. The results were compared to the reliability of the original ability scale (AB). The model computation was the same as for the baseline data, with the difference that the model parameters D, $RT_{max}$ and $AT_{max}$ obtained from the baseline analysis were used when estimating abilities at the second timepoint. Pearson correlations and Bland-Altman plots were used to compare and visualize the reliability of DT, AS and AB estimates across sessions, as well as any trend towards improved or worsened ability scores across time.

### 2.6 Age and education association

Multiple linear regression was used to compare the sensitivity of the IDoCT AS, DT estimates, and the original AB scores to age and education. Age was converted into 5-year age bins and one-hot encoded. Education was one-hot encoded into 8 categories (College or University Degree, A levels/AS levels or equivalent, O levels/GCSEs or equivalent, CSEs or equivalent, NVQ or HND or HNC or equivalent, Other professional qualifications, None of the above). Individual’s selecting “Prefer not to answer” were treated as having missing information. An alpha threshold of $p<0.01$ was used to determine significance.

### 2.7 Neural correlates of AS and DT: feature selection

A summary of the imaging association pipeline is provided in Figure 3. Models were optimised and fitted separately for each imaging modality (e.g. measures of volume, cortical thickness, fractional anisotropy and intensity) and for each set of cognitive summary scores (AS, AB and DT). First, the dataset was split into train and held-out test set with a 75/25 split. Then, age was regressed out of the imaging and cognitive feature vectors. The bivariate Spearman’s correlation between each imaging feature with the target cognitive summary score was computed for the train set only, and the features ranked into scales according to the magnitude of the obtained correlation coefficients r. Next, the combination of features that yielded the best prediction of the target variable was identified in a simple stepwise process using multiple linear regression with five-fold cross validation. Specifically, models were trained and evaluated across the five folds, iteratively removing the imaging feature with the lowest magnitude correlation coefficient *r* on the scale until just one remained within the predictor matrix. The optimal number of features was defined as the one producing models with the highest mean $R^2$ across the validation folds. The model was refitted to all of the train data using that optimal number of features. The train-test sets and folds were the same for all models, to enable cross comparison of model performance. The relative predictability of the different cognitive score estimates from the imaging data was evaluated by comparing the $R^2$ value when the optimal trained models for the different imaging modality - cognitive score combinations were applied to the held-out test data.

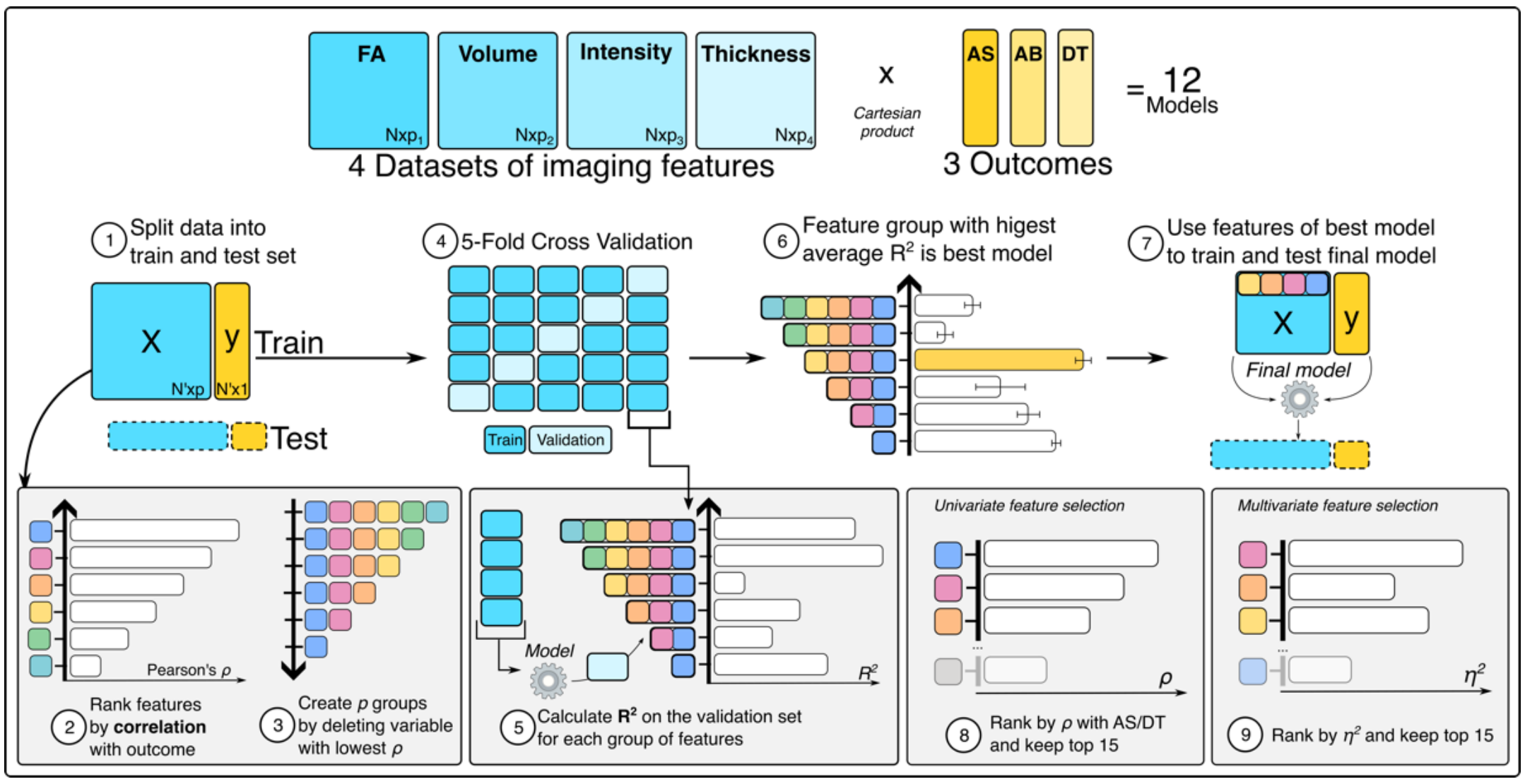

**Figure 3 Imaging analysis summary.** The imaging analysis was computed separately for each dataset (FA, Volume, Intensity and Thickness) and each potential outcome (AS, AB and DT). The analysis can be summarised in 6 main steps: 1) the data were split into train and test sets, 2) the imaging features were ranked in ascending order based on their correlation coefficient r with the outcome variable (AS, AB or DT), 3) N models were trained using 5-fold cross validation (with N equal to the number of imaging features), where in each model the feature with the next lowest coefficient r was progressively excluded from the analysis, 4) the model with the highest mean $R^2$ on the validation set was selected as the best model and the features used to train it as the best features, 5) the selected features were used to train a final multiple linear regression model using the full train test and then tested on the held-out test set, which was not used in the selection process, and 6) the eta squared of the significant features was calculated and the top 15 features with the highest eta squared as well as the top 15 features with the highest correlation coefficient r with AS or DT were identified as best neural correlates of the outcome of interest.

## 2.8 Literature review using natural language processing

The imaging features that contributed the most to the prediction of AS and DT were selected for further investigation using a Natural Language Processing (NLP) (21) pipeline in order to confirm whether they mapped onto the expected cognitive and visuomotor systems. To select features, both univariate and multivariate approaches were used. First, the $Eta^2$ values of the significant features (alpha threshold at 0.05) were calculated from the best fit models as described in the previous section. The features that were both significant and that were among the top 15 features with the highest $Eta^2$ were selected for further analysis (multivariate approach, derived from the multiple regression beta coefficients). Separately, for the univariate approach, the top 15 features with the highest magnitude of Pearson correlation with AS and DT were selected. The resultant feature labels were pooled across imaging modalities, producing AS and DT lists. We used an NLP literature search approach to summarise previously published literature in order to minimise author bias when choosing and interpreting papers related to the different brain regions.

Research papers were identified via multiple advanced search criteria based on the brain feature labels, provided in detail in the Supplementary Table A2. In general, all papers with the name of the brain region in the title and/or in the abstract, and with the words cognition and/or cognitive function in the body of the main text were selected. The addition of the "cognitive function" criteria was necessary to prevent from analysing papers that were mainly related to the cellular and biological aspects of the different brain areas. All papers that matched the advanced search criteria and that could be downloaded in HTML format from PubMed Central were included in the analysis. Papers that were only provided a PDF version, which mostly corresponded to paper published prior to 2000, were excluded, as were papers that were not open access. In total, across all brain regions, 1602 papers were analysed. Detailed numbers on how many papers were analysed for each individual brain region are provided in the Supplementary Table A2.

The HTML documents were pre-processed using the Auto-CORPus pipeline (Automated pipeline for Consistent Outputs from Research Publications) (22), an NLP tool that converts publications with an HTML structure into a BioC JSON format (23). The BioC JSON format uses a standard structure developed to allow for the interoperability of text mining outputs across different systems. Concretely, it consists of collections of documents, extracted from a corpus, characterized by different elements, that contain the actual text as well as additional information about the original document. The text is automatically divided into common sections (e.g. abstracts, results ...) and paragraphs, and each section is associated to the respective, and unique, Information Artifact Ontology (IAO) annotation (24). IAO annotations enable identification of the same sections across different publications, even when they are associated to different titles (e.g. Methods and methodology). Specifically, for our NLP analysis, only the sections corresponding to the Abstract, Discussion, and Introduction were used, which corresponded respectively to the IAO annotations IAO:0000316, IAO:0000315, and IAO:0000319.

The extracted paragraphs were cleaned by converting all words to lower case, removing words that were less than 2 characters or more than 20 characters long, by removing all special characters and punctuations, and by removing all stop words, which correspond to commonly used words in the English language, such as "you", "an", or "in". In addition, words that are related to the different brain structures (e.g. gyrus, nucleus, cerebellum…), or that are commonly associated to the study design (e.g. controls, patients, humans …) were excluded. After the data cleaning, the frequency of occurrence of each word was calculated across all papers for each individual brain region, and normalized between 0 and 1, with 1 corresponding to the most frequent word.

In order to assess the main brain functions associated to AS and DT, the frequency scores of the 5 words with the highest values were added separately for all the brain regions related to AS and DT. In this way, if a word appeared frequently in multiple brain regions, then it was assigned a higher frequency score, compared to a frequent word appearing in one individual brain region.

#### 2.9 Software

The analysis was conducted in Python (3.7.1). The main modules used were pandas (1.3.5), numPy (1.21.5), pingouin (0.5.3), statsmodels (0.13.1), nltk (3.7), and genism (4.2.0). The visualization was completed with seaborn (0.11.0), matplotlib (3.5.1) and wordcloud (1.8.2.2) in Python and ggplot2 (3.3.6) in R (4.0.1).

## 3. RESULTS

#### 3.1 Samples characteristics

The full sample consisted of 34927 participants at baseline, among which only 33890 fully completed the PVT. The mean age was $64.7 \pm 7.8$ and 53% of the participants had an education level comparable to A/AS levels, or higher. Among these 34927 participants, 4962 people (mean age: $61.7 \pm 7.2$) completed the PVT at a follow up timepoint. Full details on the sample demographics are available in Table 1.

**Table 1 Sample demographics.** Age and education of the participants of the study at baseline and at the follow up timepoint

| Variable | Category | Number (percentage) | |
|---|---|---|---|
| | | Baseline | Follow up |
| **TOTAL COUNT** | | 34927 | 4962 |
| Age group (years) | 40-50 | 346 (1%) | 54 (1%) |
| | 50-60 | 9525 (27%) | 1557 (39%) |
| | 60-70 | 14225 (41%) | 1661 (42%) |
| | 70-80 | 10451 (30%) | 661 (17%) |
| | >= 80 | 380 (1%) | 6 (0%) |
| | Missing information | 0 (0%) | 0 (0%) |
| Educational level | College or University Degree | 16435 (47%) | 1888 (48%) |
| | A levels/AS levels or equivalent | 12828 (37%) | 1519 (39%) |
| | O levels/GCSEs or equivalent | 18567 (53%) | 2174 (55%) |
| | CSEs or equivalent | 4649 (13%) | 633 (16%) |
| | NVQ or HND or HNC or equivalent | 6558 (19%) | 802 (20%) |
| | Other professional qualifications (eg: nursing, teaching) | 12783 (36%) | 1447 (37%) |
| | None of the above | 2034 (6%) | 165 (4%) |
| | Prefer not to answer | 116 (0.3%) | 54 (1%) |
| | Missing information | 419 (1%) | 0 (0%) |

### 3.2 Measures of individual trial difficulty, ability and visuo-motor speed from IDoCT

The IDoCT model converged after 250 iterations when estimating trial difficulty measures (D), reaching a mean percent change in word difficulty that tended to 0. In the second iteration, necessary to determine AS and DT, the model converged in 10 iterations while defining a measure of ability and performance. DS was calculated in a final step of the model computation.

The distributions of D (unscaled difficulty) and DS (scaled difficulty) are available in Figure 4A and 4B, together with the original difficulty scale ($D_{old}$) (Figure 4C), and the association between DS and $D_{old}$ (Figure 4D). The mean D, DS and $D_{old}$ were respectively $0.60 \pm 0.07$, $0.72 \pm 0.06$, and $0.51 \pm 0.24$. The full list of the words presented and of their associated measure of D, DS and $D_{old}$ is available in the Supplementary Table A3. In brief, according to the IDoCT D scale, the five most difficult words, in ascending order of difficulty, were: glower, malefactor, pachyderm, matron and plethora. Instead, the five easiest words in ascending order of difficulty were: calm, weld, herd, desolate and engraved. On the other hand, according to DS, the top 5 hardest and easiest words were respectively buffet, trivet, prodigious, bucolic, truncate, and fabricate, angry, monarch, fly, run. In general, most of the words that were assigned to higher difficulty scores were of Latin, Greek or French origin.

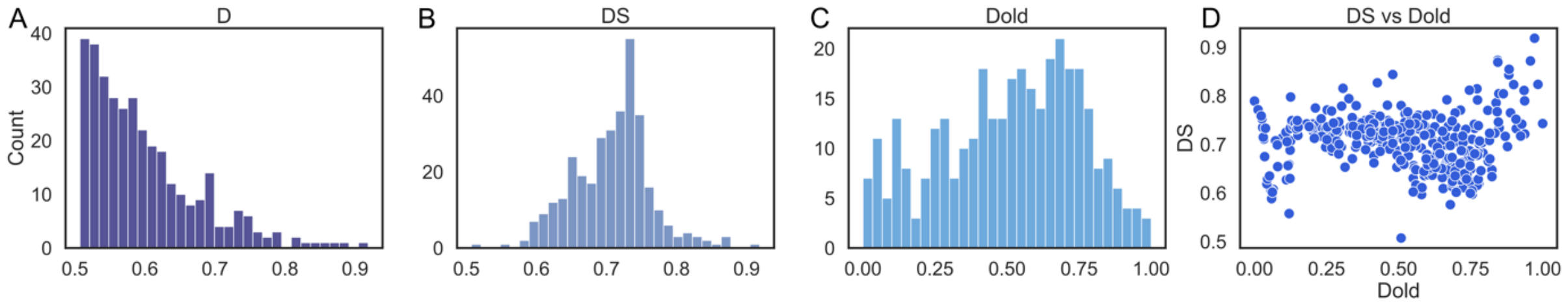

**Figure 4 D, DS and Dold distributions,.** Comparison of the distributions of IDoCT measures of A) unscaled trail's difficulty D, 2) scaled trial's difficulty D and C) original difficulty Dold

The change in assigned difficulty level per word-picture pair between D and DS is presented in Figure 5. In brief, the difficulty of words that were presented exclusively to participants with lower abilities tended to decrease after the scaling. On the other hand, words that were assigned only to participants with higher abilities tended to increase in difficulty after scaling. This pattern of results accords with the method working to correct for sampling bias.

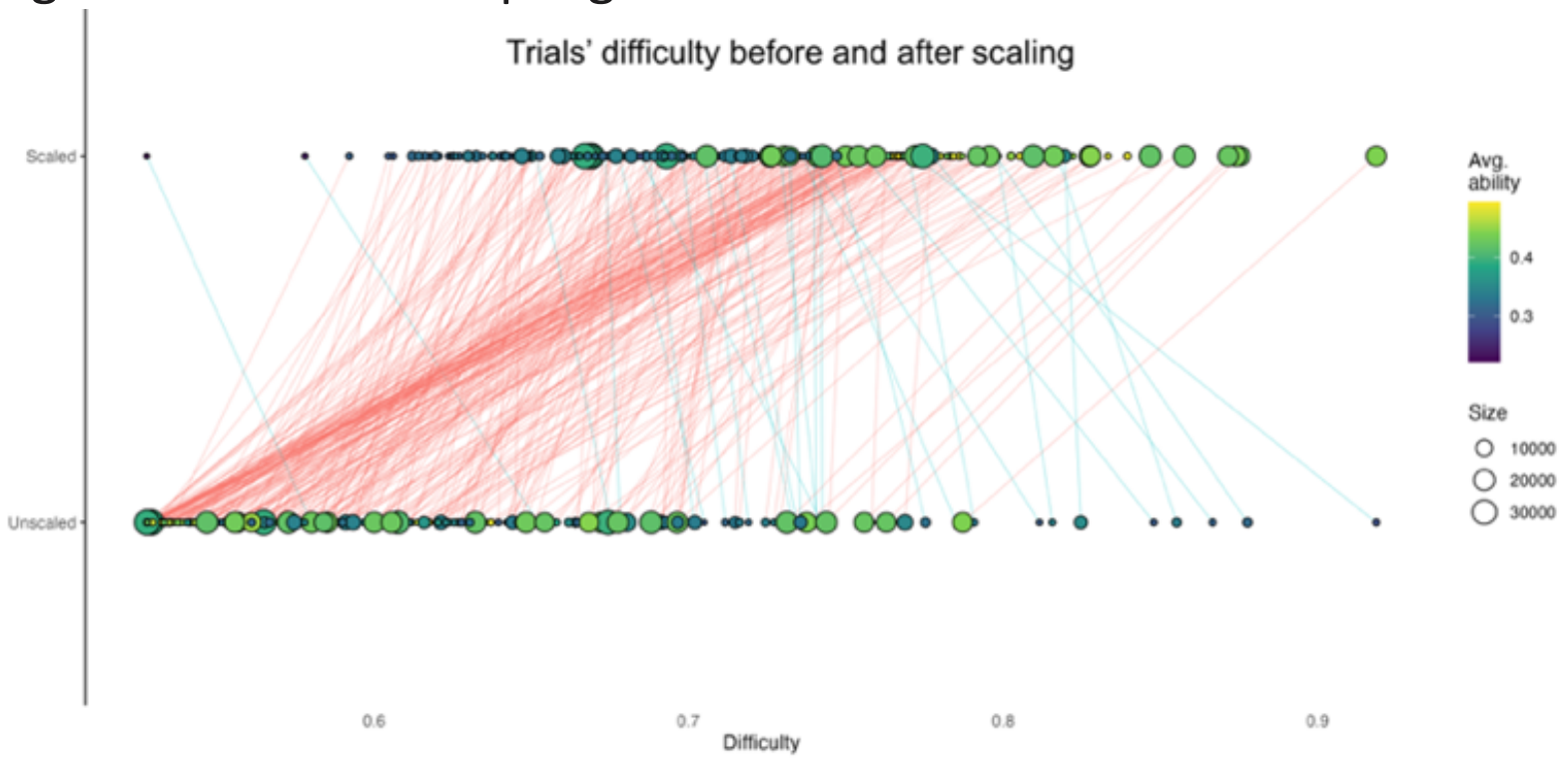

**Figure 5 Change in word difficulty IDoCT scores before and after scaling.** The data-driven measures of trial's difficulty D were scaled according to the ability of the participants to which the trials were presented, in order to account for potential biased sampling in the task design, meaning cases in which words were presented exclusively to participants with high/low ability. In the figure, a red and blue line correspond respectively to the increase and decrease in difficulty score for each word after the scaling. Each dot in the figure corresponds to one word, the size of the dot indicates the number of participants that were presented that word (i.e. bigger dots mean that a word was presented to a higher number of participants) and the color is related to the average AS of the participants that were presented a specific word (or trial).

The mean AS predicted by IDoCT across the cohort was 0.39 $\pm$ 0.08, while the mean DT was 3160 $\pm$ 1081. The mean AB was 0.85 $\pm$ 0.09. The distributions of AS, DT and AB were compared, as well as their associations to the raw median RT and the number of correct answers, as presented in Figure 6. As can be observed, AB was characterized by an atypical bimodal distribution centered around high ability scores (~0.85), which supports the hypothesis of the ceiling effect. On the other hand, both DT and AS had the expected near-gaussian distribution of abilities, which, in case of AS, was centered around average scores (~0.4).

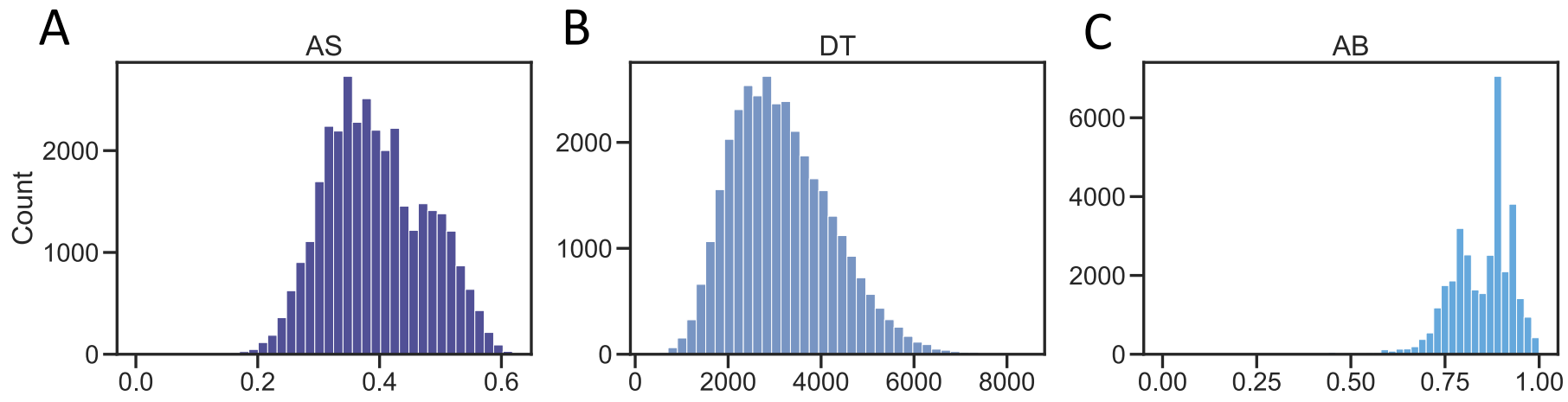

**Figure 6 A, AS and AB distributions at baseline.** Comparison of the distributions of IDoCT measures of A) AS (cognitive ability), 2) visuo-motor speed DT and C) original ability measure AB.

By comparing the associations of AB and AS (Figure 7) with the number of correct replies and the median RT, expected associations are observed in case of AS, with more correct answers and faster

RTs being related to higher AS. On the other hand, in case of AB, some participants were assigned a low AB score despite giving correct answers in the majority of cases and despite having low median RTs.

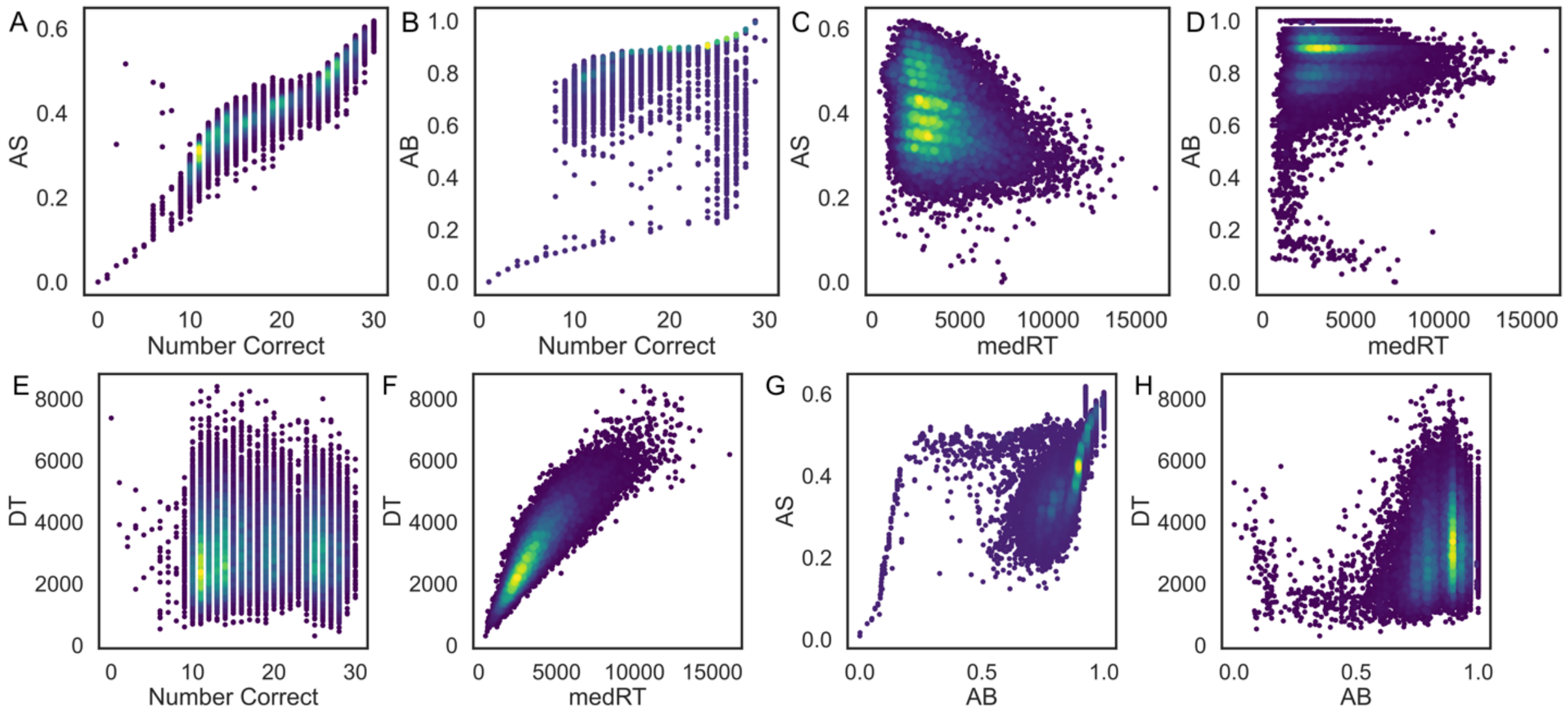

**Figure 7 AS and AB association with median RT and number of correct answers.** A) Association between AS and number of correct answers, B) Association between AB and number of correct answers, C) Association between AS and median RT, and D) Association between AB and median RT, E) Association between DT and number of correct answers, F) Association between DT and median RT, G) Association between AS and AB, and H) Association between DT and AB

In case of DT, no association with the number of correct answers was observed (Figure 7E), which is expected considered that the level of visuo-motor latency should not affect how correctly each participant replies. On the other hand, DT was almost linearly associated with the median RT (Figure 7F), with higher median RTs for participants with longer DT. This is also expected since participants with longer visuo-motor latency are expected to be overall slower. No clear association was observed between AS and AB and DT and AB. The latter is not surprising considered that DT is not supposed to capture cognitive performance like AB. On the other hand, the association between AS and AB resembles that of AB and the number of correct, which is expected considered that AS is almost linearly associated with accuracy (Figure 7G and 7H).

### 3.3 Trials-by-trial difficulty trajectories

To validate that $D_{old}$ led to ceiling effects, trial difficulty trajectories were plotted, that is, showing how the difficulty of sampled word-picture combinations changed sequentially through the task. Separate trajectories were computed for D, DS and $D_{old}$, and for participants with different ability levels. Participants were divided into 10 groups based on AB (0-0.1, 0.1-0.2, … 0.9-1.0), and their mean trajectories obtained by averaging the difficulty scores of the words presented at each, as shown in Figure 8. If the adaptive staircase approach was correct, then a gradual increase in difficulty should have been observed for all participants with medium-high ability (AB > 0.3). Instead, already after up to 5 trials, $D_{old}$ either reached a plateau or started to decrease, which suggests that no words with higher difficulty as defined by $D_{old}$ were available after that point, forcing the algorithm to provide words of either equal or lower difficulty. Furthermore, the difficulty level as defined by D and DS appeared not to increase across time. Together, these results support the hypothesis that the original difficulty scale was not optimally calibrated for the assessed cohort.

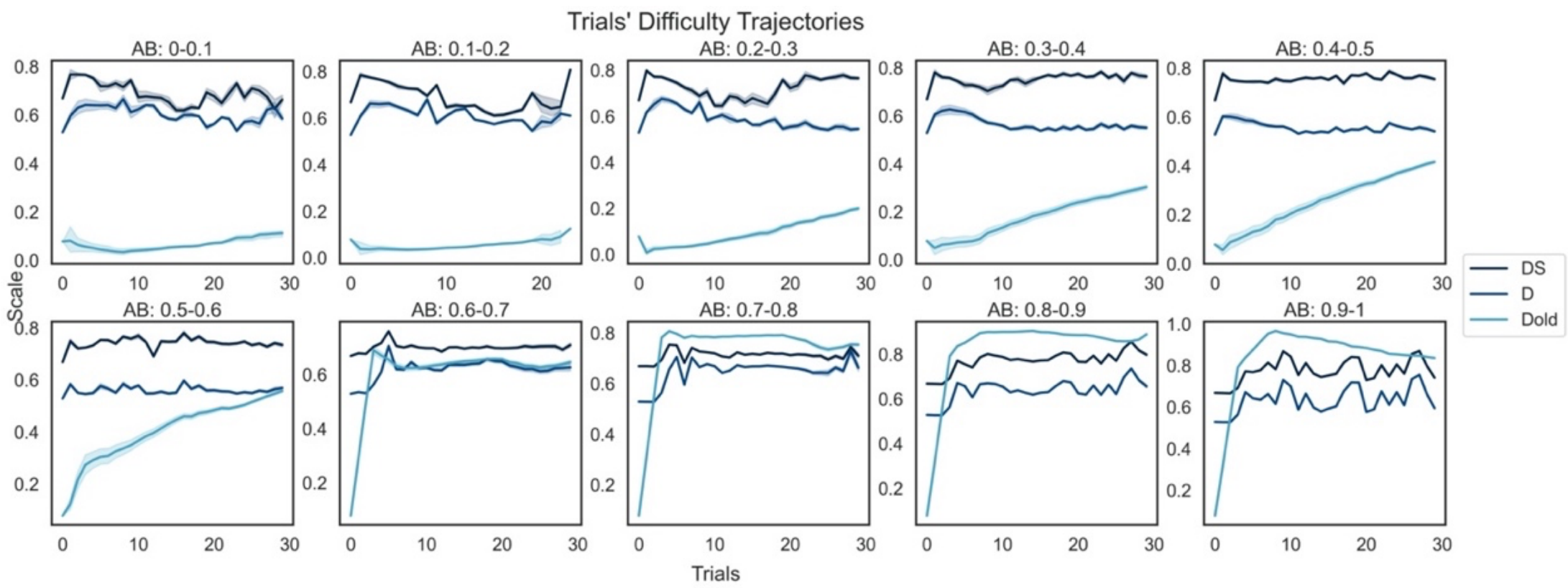

**Figure 8 Trial's difficulty trajectories**. Participants were divided into 10 groups based on their AB measures. The trial difficulty trajectory of each group was obtained by measuring the mean difficulty (D, DS and $D_{old}$) of the presented trial's at each step of the cognitive assessment. In case of participants with medium-high difficulty, $D_{old}$ either reached a plateau or decreased after few trials, meaning that the original difficulty scale was not optimally calibrated.

## 3.4 AS has better test-retest reliability compared to AB

IDoCT was applied to the follow up data to evaluate retest reliability. The model converged after 250 and 10 iterations respectively when predicting D and AS/DT. The distributions of the predicted AS and DT, as well as of AB, are available in Figure 9. The mean AS, DT and AB at follow up were respectively $0.4 \pm 0.08$, $3009 \pm 1035$ and $0.83 \pm 0.09$. Similar to the baseline, both AS and DT were characterised by gaussian shaped distributions, with AS being centered around average values (~0.4), while AB again consisted of a bimodal distribution centered around high ability scores.

By comparing the iDoCT predictions at baseline and follow up, both AS and DT had a good test-retest reliability (r = 0.77 for AS, and r = 0.57 for DT) across all ranges of scores (Figure 10). The test-retest reliability of AB was good but substantially lower compared to AS (r = 0.66). Furthermore, the distribution of the Bland-Altman plots showed a poor spread and only high AB scores being consistent across timepoints.

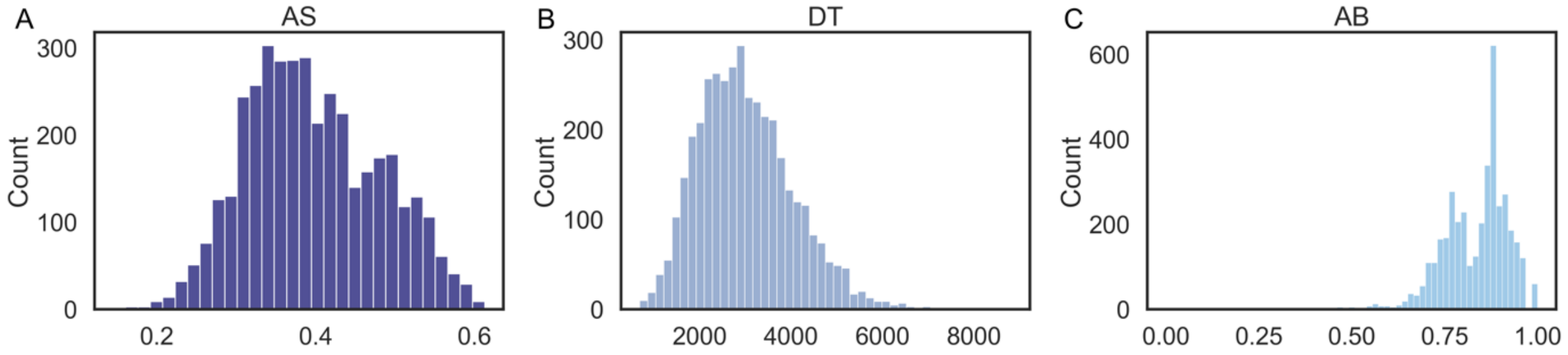

**Figure 9 A, AS and AB distributions at follow up.** Comparison of the distributions of IDoCT measures of A) AS (cognitive ability), 2) visuo-motor speed DT and C) original ability measure AB.

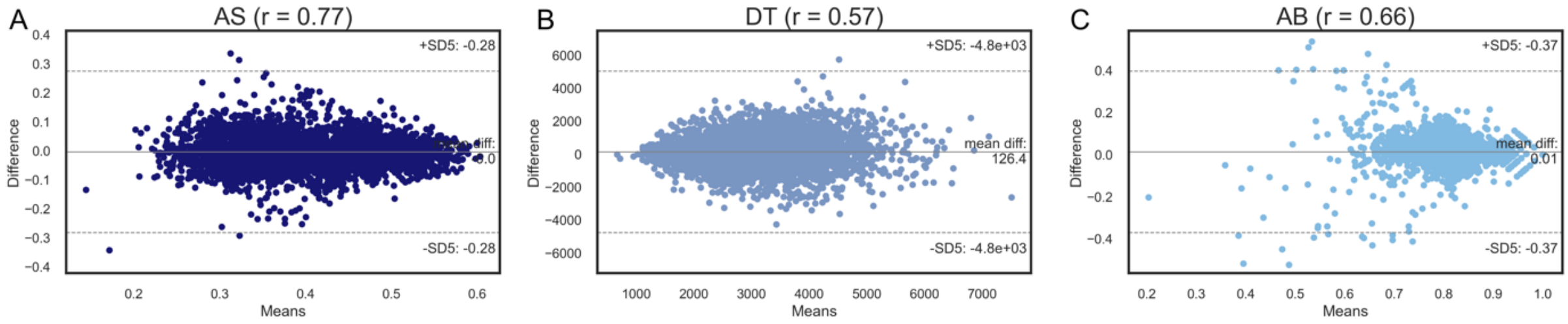

**Figure 10 Bland-Altman plots of DT, AS and AB.** The correlation coefficients r between AS (A), DT (B) and AB (C) at baseline and follow up were respectively 0.77, 0.57 and 0.66.

### 3.4 Older and better educated participants have higher cognitive abilities scores

When predicting AS from age decade and education level, a significant regression equation was found (F = 710.6 (13, 33876), p = < 0.001) with an $R^2$ of 0.21. The regression equation for the original AB score was also significant (F = 561.2 (13, 33864), p = < 0.001) but with a lower $R^2$ of 0.17. Both age and education were significant predictors of AS and AB. Regarding education, the strongest predictors were having a college/University degree (ed1) and A/AS level (ed2), which had respectively a positive effect size in standard deviation (SD) units of 0.66 and 0.35 for AS, and somewhat less at 0.57 and 0.33 SDs for AB, compared to the reference category. In the case of age, older participants had higher AS compared to participants aged 45. The increase was gradual until age 70, at which point a plateau was reached, and participants reached a 0.48 SD increase in AS vs 0.41 increase in AB (for participants aged 75 compared to the reference category) (Figure 11).

### 3.5 Older participants have longer visuo-motor latency times

In the case of DT, a significant regression equation was found (F = 39.41 (10, 33876), p = < 0.001) with an $R^2$ of 0.015. Both age and education were significant predictors of DT, but the effect size of education was consistently small to negligible (below 0.1 SD units). The relationship between age and DT was comparable to the association with AS, with older participants having a gradual increase in DT, or visuo-motor latency time. More specifically, participants at age 75 and 80 were associated to respectively a 0.38 SD and 0.52 SD increase in DT compared to participants at age 45 (Figure 11).

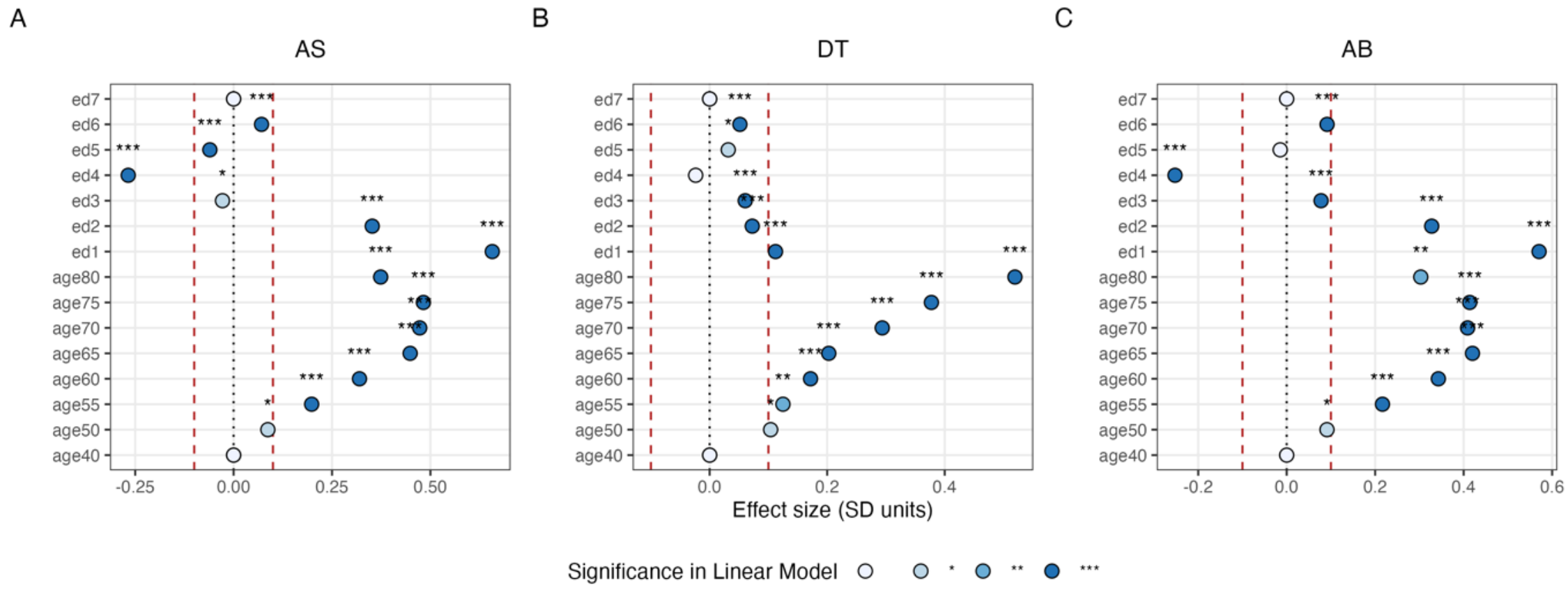

**Figure 11 Analysis of age and education associations.** Multiple linear regression models were trained using AS (A), DT (B), and AB (C) as predicted variables and age and education as predictors. The obtained beta coefficients in SD units are represented in the figure. Age and education were always significant predictors, but the effect size of education was negligible (-0.1 < x > 0.1) in case of DT. The color and asterisk represent the significance level of each feature. Age45 and ed7 were used as reference category. ed1 corresponds to College/University degree (ed1), ed2 to A levels/AS levels or equivalent, ed3 to O levels/GCSEs or equivalent, ed4 to, CSEs or equivalent, ed5 to NVQ or HND or HNC or equivalent, ed6 to other professional qualification, and ed7 to none of the above. The red lines indicate the area where the effect size can be considered negligible.

### 3.6 Imaging analysis feature selection

The results of the fine-tuning are reported in Figure 12 for AS and Figure 13 for DT. In total, P models were trained for each dataset (i.e. FA, Thickness, Volume and Intensity), where P corresponds to the number of features available. Model0 corresponds to the model trained on all features except for the feature with the lowest correlation coefficient. For each of the following models, the feature with the lowest coefficient r among those remaining was dropped, until Model(P-2), which was trained exclusively on the last feature available (i.e. the feature with the highest r correlation coefficient).

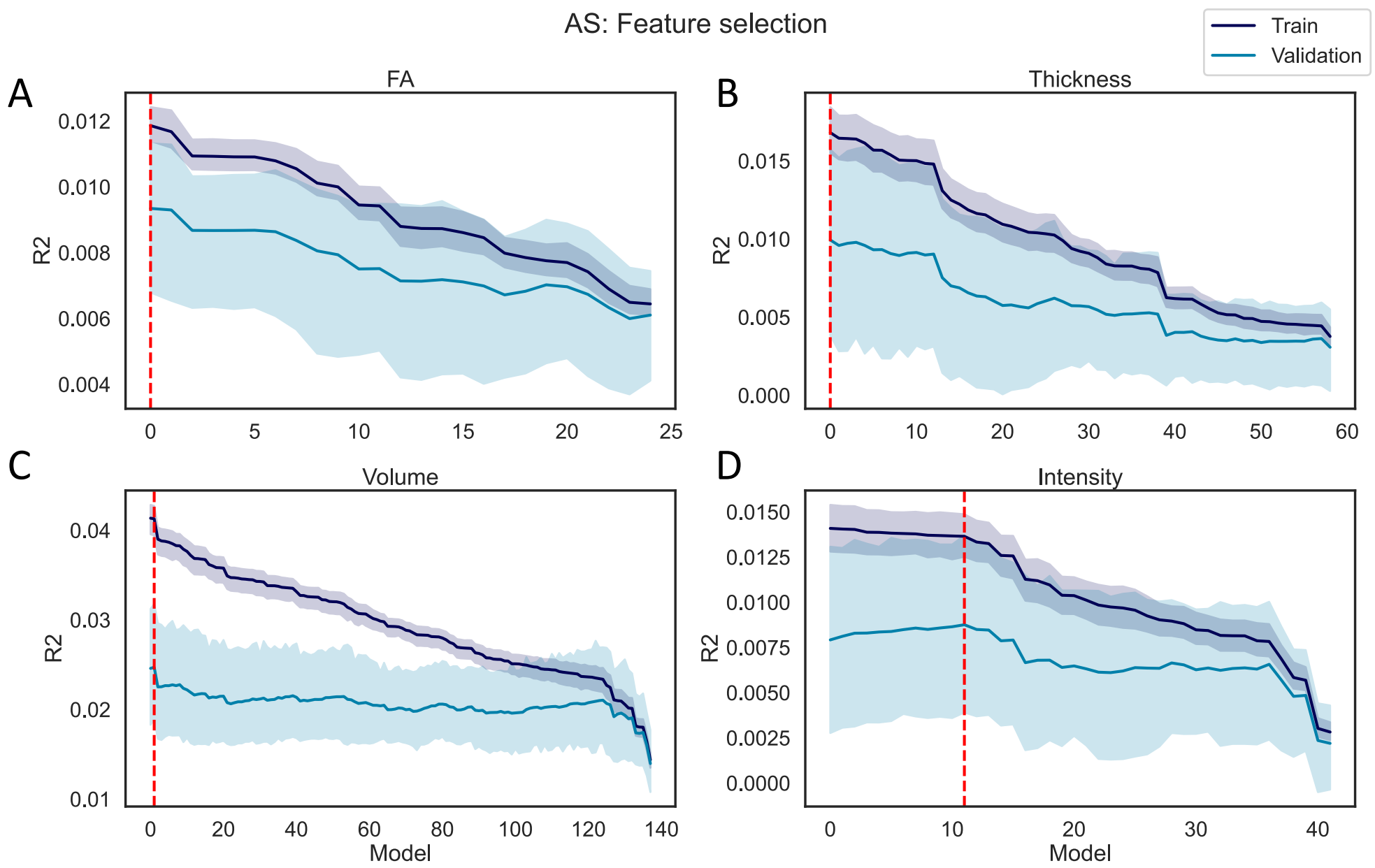

**Figure 12 AS feature selection.** P models were trained using as regressors the imaging data in each individual dataset (A) FA, B) Thickness, C) Volume and D) Intensity) and as predictor AS, where P corresponds to the number of features in each dataset. In each iteration of the model, the features (among those still available) with the lowest correlation coefficient with the predictor was excluded from the regressors, meaning that Model1 was trained on all features except 1 (i.e. the feature with the lowest r), and ModelP was trained only on one feature (i.e. feature with highest r). The $R^2$ was calculated on the train and validation set. The train/validation $R^2$ (and their standard deviation across 5 folds) are reported in the figure. The red line represents the best performing model, namely the model that yielded the highest $R^2$ on the validation set.

When predicting AS, in the case of FA and Thickness, Model0 was the best performing, which included all features except for respectively the FA in the right cingulate gyrus (part of cingulum), and the thickness in the right pars triangularis. Similarly, Model1 was the best performing for Volume, meaning that only two features (i.e. the Brain Stem and Crus I Cerebellum vermis) were dropped when predicting AS. Finally, in case of Intensity, Model11 was the one that yielded the highest $R^2$ in the validation set, which resulted in 12 features being dropped (i.e. mean intensity of CSF, White Matter hypointensities, non-White matter hypointensities, Corpus Callosum Mid-Posterior, $5^{th}$ Ventricle, left/right Cerebellum White Matter, Corpus Callosum Posterior, right/left Cerebellum Cortex, left Amygdala and volume of White Matter hypointensities). The $R^2$ obtained for each model, and each dataset, is available in the Supplementary Tables A4-A7.

When predicting DT, Model12 was the best FA model, which resulted from dropping 13 features (i.e. FA in the left/right superior thalamic radiation, left/right corticospinal tract, right posterior thalamic radiation, left inferior longitudinal fasciculus, left superior longitudinal fasciculus, left parahippocampal part of cingulum, middle cerebellar peduncle, right/left inferior fronto occipital fasciculus, left anterior thalamic radiation and forceps minor). For Volume and Thickness, Model97 and Model45 were the best performing, as a result respectively only 41 out of 139 and 13 out of 60 features were kept. When fine-tuning the models on the intensity dataset, all the models had $R^2$ <=0. The only model with a positive $R^2$ ($R^2$ = 0.001) was obtained after dropping all the features except one. Due to the low performance of the intensity-based models, no features were extracted from the intensity dataset during the feature selection step. More detailed information on which features were dropped during the fine-tuning is available in the Supplementary Tables A8-A11, as well as the obtained $R^2$ on the train and validation set of each model.

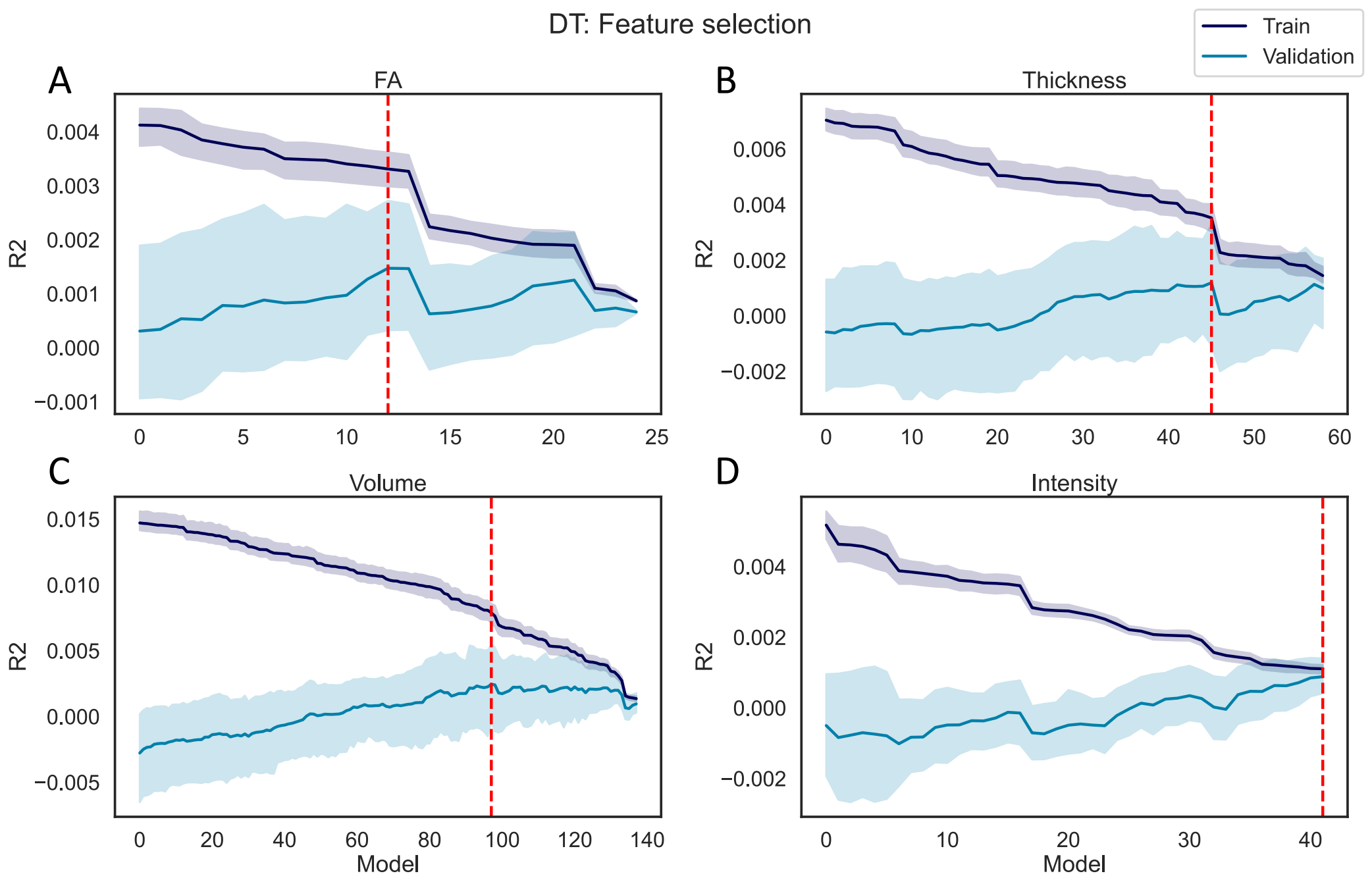

**Figure 13 DT feature selection.** P models were trained using as regressors the imaging data in each individual dataset (A) FA, B) Thickness, C) Volume and D) Intensity) and as predictor DT, where P corresponds to the number of features in each dataset. In each iteration of the model, the features (among those still available) with the lowest correlation coefficient with the predictor was excluded from the regressors, meaning that Model1 was trained on all features except 1 (i.e. the feature with the lowest r), and Model(P-2) was trained only on one feature (i.e. feature with highest r). The $R^2$ was calculated on the train and validation set. The train/validation $R^2$ (and their standard deviation across 5 folds) are reported in the figure. The red line represents the best performing model, namely the model that yielded the highest $R^2$ on the validation set.

Results of the feature selection process when applying the same train-validation pipeline to AB are reported in Figure 14.

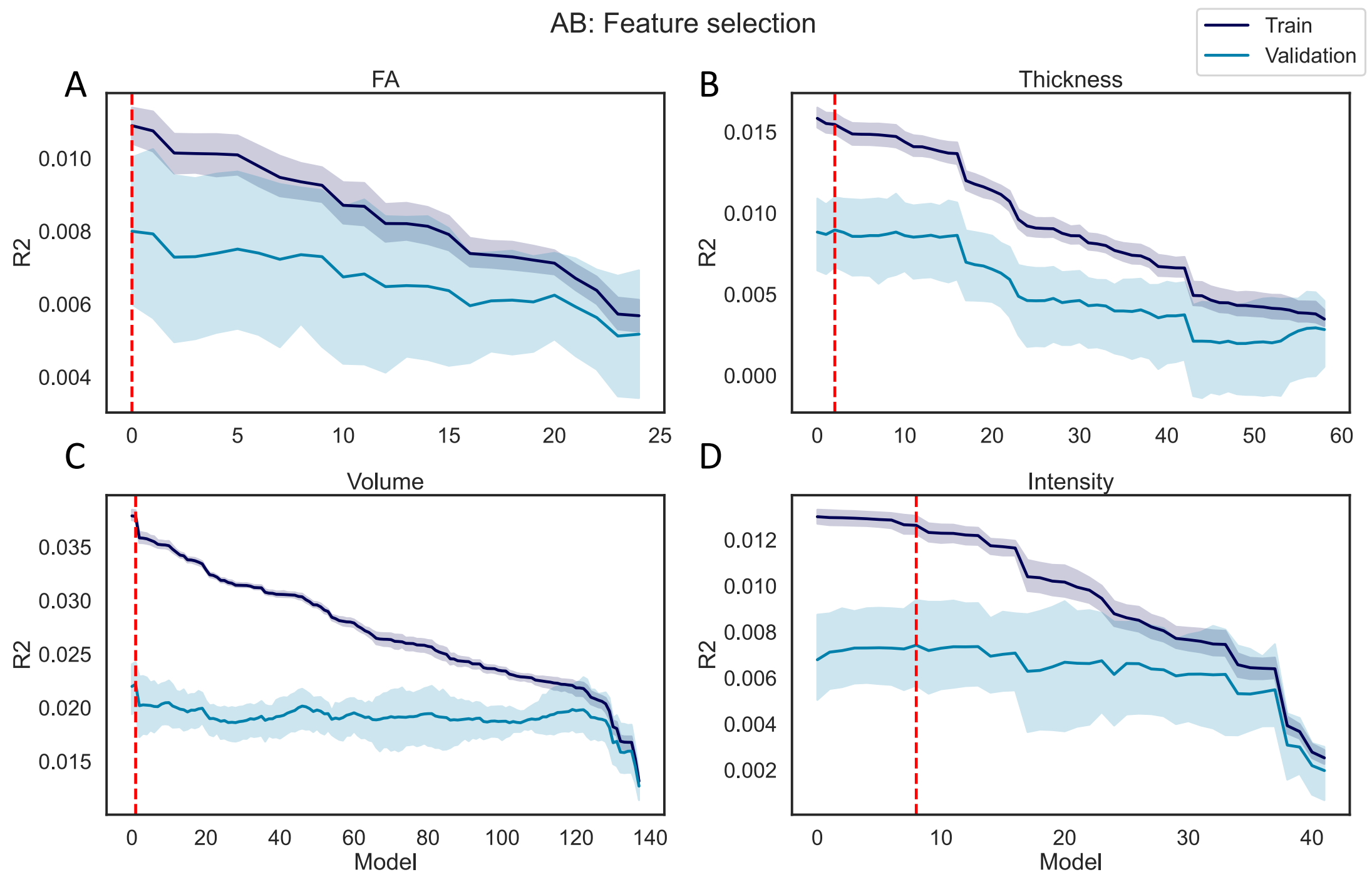

**Figure 14 AB feature selection.** P models were trained using as regressors the imaging data in each individual dataset (A) FA, B) Thickness, C) Volume and D) Intensity) and as predictor AB, where P corresponds to the number of features in each dataset. In each iteration of the model, the features (among those still available) with the lowest correlation coefficient with the predictor was excluded from the regressors, meaning that Model1 was trained on all features except 1 (i.e. the feature

with the lowest r), and Model(P-2) was trained only on one feature (i.e. feature with highest r). The $R^2$ was calculated on the train and validation set. The train/validation $R^2$ (and their standard deviation across 5 folds) are reported in the figure. The red line represents the best performing model, namely the model that yielded the highest $R^2$ on the validation set.

**Table 2 Multiple linear regression models with imaging features.** The features identified during the feature selection step were used as regressors of multiple linear regression models to predict AS, AB and DT. The $R^2$ (train and test), F statistics and p-value of the models are reported in the table.

| | AS | | | | AB | | | | DT | | | |
|---|---|---|---|---|---|---|---|---|---|---|---|---|
| **Dataset** | $R^2_{train}$ | $R^2_{test}$ | **p-value** | **F** $_{train}$ | $R^2_{train}$ | $R^2_{test}$ | **p-value** | **F (df)** $_{train}$ | $R^2_{train}$ | $R^2_{test}$ | **p-value** | **F (df)** $_{train}$ |
| FA | 0.011 | 0.016 | <0.001 | 9.539 | 0.011 | 0.014 | <0.001 | 8.738 | 0.003 | 0.003 | <0.001 | 4.924 |
| Intensity | 0.013 | 0.008 | <0.001 | 8.651 | 0.012 | 0.006 | <0.001 | 7.247 | - | - | - | - |
| Grey volume | 0.037 | 0.023 | <0.001 | 5.800 | 0.034 | 0.019 | <0.001 | 5.308 | 0.007 | 0.001 | <0.001 | 3.594 |
| Thickness | 0.016 | 0.015 | <0.001 | 5.588 | 0.015 | 0.012 | <0.001 | 5.321 | 0.002 | 0.001 | <0.001 | 3.056 |

### 3.7 Identification of neural correlates of AS and DT: univariate and multivariate analysis

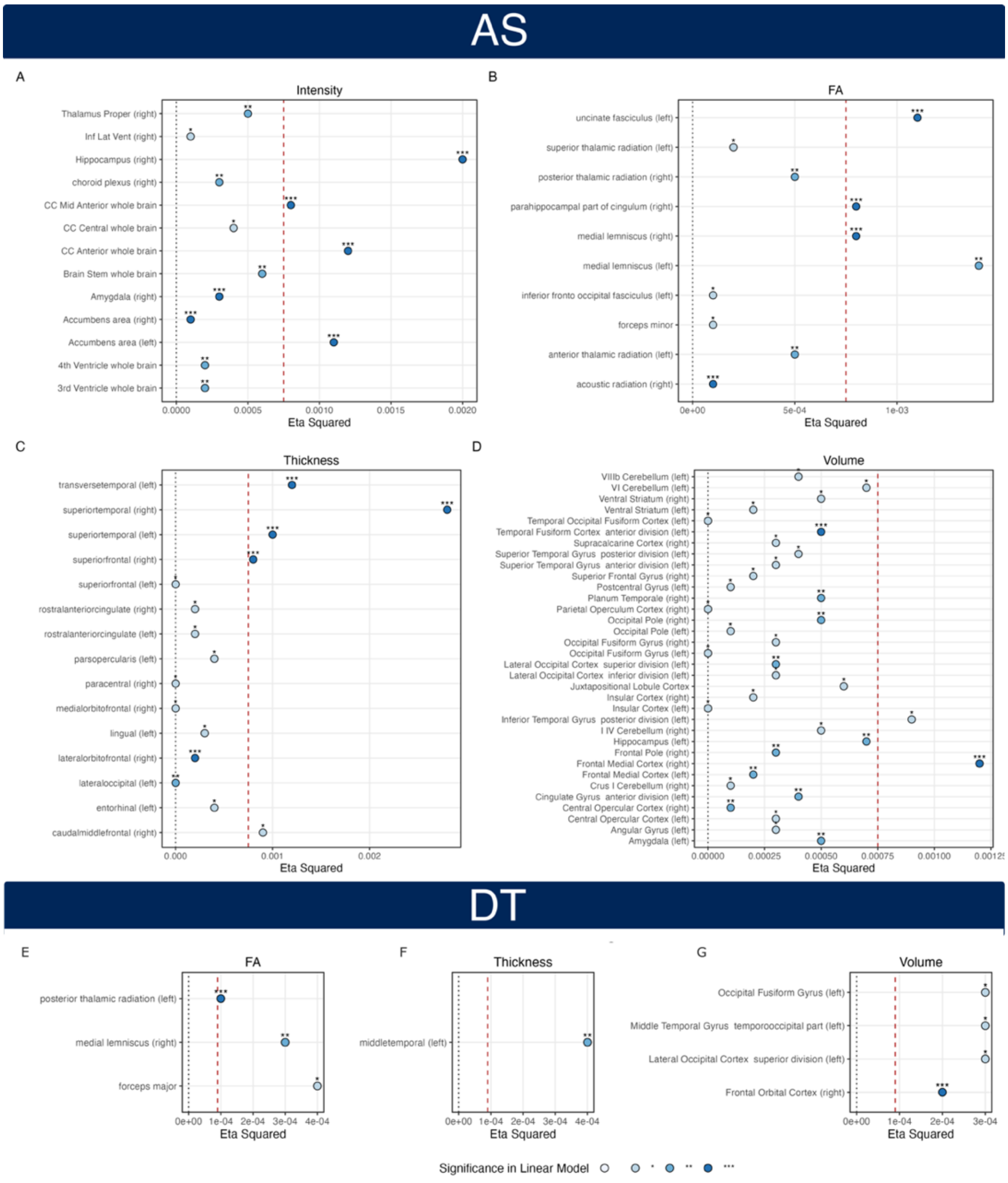

**Figure 15 Multiple linear regression using imaging features.** The imaging features identified during the feature selection step were used as regressors of multiple linear regression models to predict AS (A, B, C, D) and DT (E, F, G). The measured eta squared of the features identified as significant in the models are reported in the figure. The color and asterisk represent the significance level of each feature. The red dotted line separates the features belonging to the top 15, which were selected as a neural correlate of AS/DT, from the other features.

The selected features were used as regressors of multiple linear regression models trained on the full train set and tested on the held-out test set. Two models were trained for each dataset, predicting either AS, or DT. Overall, significant regression models were found for all analyzed datasets, with an average train and test $R^2$ that was respectively 0.02 $\pm$ 0.01 and 0.016 $\pm$ 0.005 across all datasets for AS, and 0.004 $\pm$ 0.002 and 0.002 $\pm$ 0.001 across all datasets for DT. Full results are available in Table 2. The model for the intensity dataset and DT was not computed due to the identified low performance at the feature selection step of the analysis. For AB, significant regression models were found for all analyzed datasets, with an average train and test $R^2$ that was respectively 0.018 $\pm$ 0.01 and 0.013 $\pm$ 0.005. Full results are available in Table 2. Therefore, while the models predicting AS had modest R2 values, they performed numerically better than the models predicting AB across all four imaging datasets.

The significant features of the AS and DT models, and their respective $Eta^2$ are presented in Figure 15. The 15 top significant features with the highest $Eta^2$ across all data modalities were identified as the best neural correlates of AS and DT and used in the next steps of the analysis. The identified neural correlates of AS were: Hippocampus, superior and transverse temporal gyrus, medial lemniscus, anterior corpus callosum, medial frontal cortex, uncinate fasciculus, nucleus Accumbens, caudal middle frontal gyrus, inferior temporal gyrus, superior frontal gyrus, and parahippocampus part of cingulum. The identified neural correlates of DT were forceps major, medial lemniscus, posterior thalamic radiation, middle temporal gyrus, lateral orbitofrontal cortex, occipital fusiform gyrus, and lateral occipital cortex.

In order to assess the robustness of the results, a second set of neural correlates were identified by completing a univariate analysis. In this case, the selected features were ranked according to the magnitude of their Pearson correlation coefficient obtained after correlating each feature with either AS or DT. The top 15 features across all data modalities with a significant p-value (p-value < 0.001) were selected as neural correlates of AS and DT. The identified neural correlates of AS were: Amygdala, Hippocampus, Frontal Pole, Insular Cortex, Cerebellum, Superior temporal gyrus, and Temporal fusiform cortex. On the other hand, the identified neural correlates of DT were: middle temporal area, Cerebellum, Inferior temporal gyrus, Intracalcarine cortex, Occipital Pole, superior temporal and lateral orbito-frontal area.

### 3.8 DT is associated more strongly to brain regions with visuo-motor functions and AS to regions with memory and language functions

The full list of the words with the highest frequency of occurrence in the articles that use the anatomical labels associated with AS and DT is available in the Supplementary Tables A12-A15, both for the univariate and multivariate analysis. A summary of the results is presented in Figure 16.

In summary, Visual (2.57), Age (2.0), motor (1.8) lesion (1.32) and Stimulus (1.27) were the words with the highest frequency of occurrence for DT associated anatomical labels in case of the multivariate analysis. Similarly, visual (3.63), stimulus (1.53), and lesion (1.29) were the ones with the highest frequency for DT associated anatomical labels derived from the univariate analysis. Conversely, Memory (3.75), Age (3.24), and Auditory (2.0), and Memory (2.0), social (1.68), behavior (1.59), emotional (1.33) and learning (1.28) were the most frequent words in case of AS (Figure 16) for respectively the multivariate and univariate analysis.

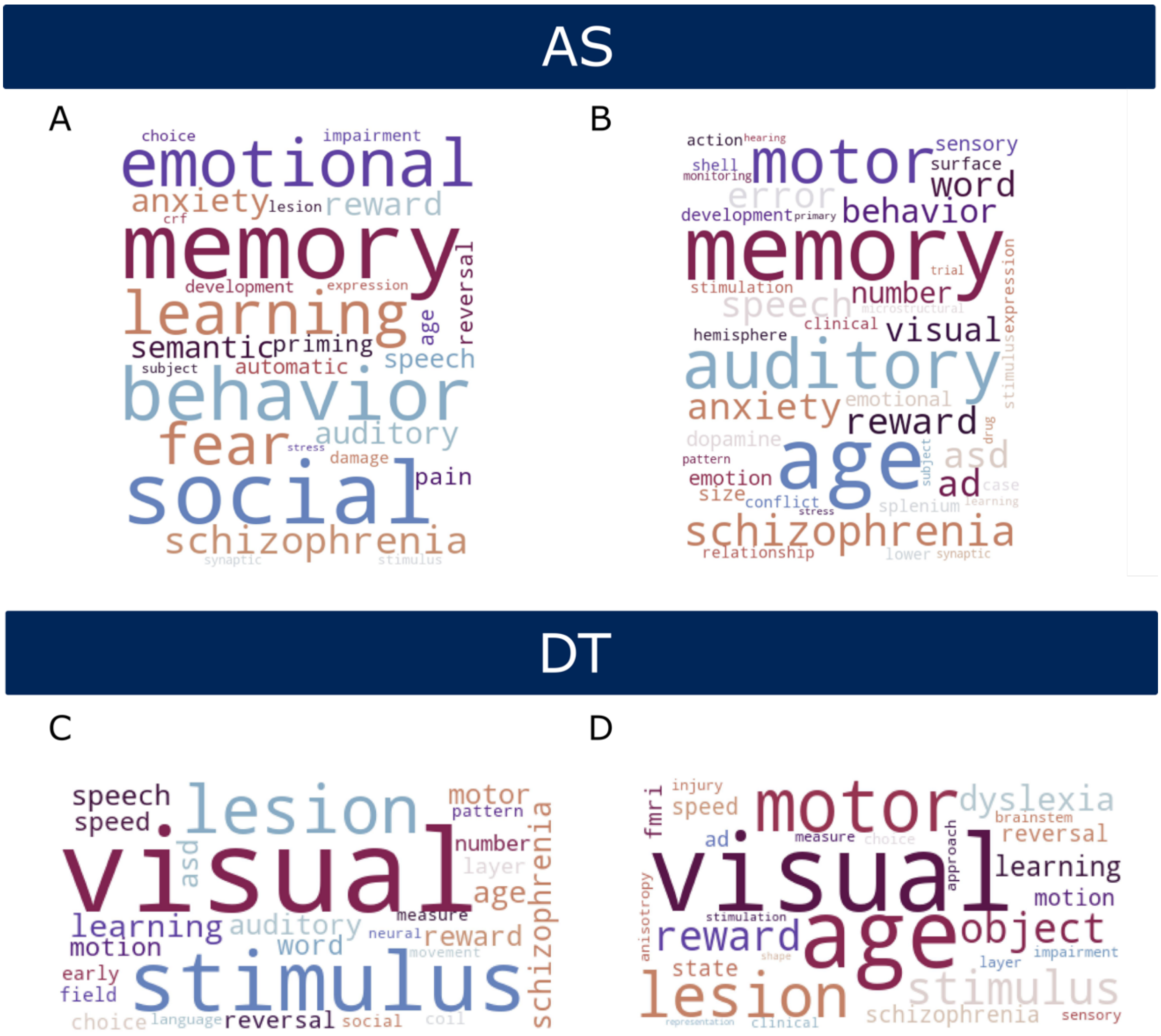

**Figure 16 NLP analysis results.** The word clouds represent the words with highest frequency of occurrence in papers of the best neural correlates of A) AS derived from univariate analysis, B) AS derived from multivariate analysis, C) DT obtained from univariate analysis, and D) DT obtained from multivariate analysis. Bigger words correspond to higher frequency of occurrence.

## 4. DISCUSSION

IDoCT is a flexible method designed to fractionate the detailed timecourses that are collected during performance of a computerised cognitive task into components that can be explained by inter-subject variability in basic visuomotor processing speed and device latency on the one hand, and the specific cognitive abilities that the task was intended to manipulate on the other (14). This is achieved in a simple, robust, and data-driven manner that iteratively re-estimates individuals' abilities and trial-difficulty scales whilst handling the speed accuracy tradeoff.

This method was initially applied to improve the precision of performance estimates from the Cognitron library of online tasks (14). However, the flexibility of the approach can be adapted for practically any computerised task that varies dimensions of cognitive difficulty across trials and where performance recordings are available for large numbers of individuals. The results presented here provide further evidence of the utility of IDoCT in the context of data derived from an independently designed UK Biobank task, including improvements in the trial-difficulty scale, participants-score distributions, retest statistics, demographic correlations, and imaging associations.

More specifically, a critical limitation of the PVT dataset is that the original summary score distributions are malformed due to the dynamic sampling algorithm having operated across a sub-optimal trial-difficulty scale ($D_{old}$). Our analysis of sampling trajectories highlights the basis of that limitation. Specifically, the difficulty ($D_{old}$) of sampled trials for participants with moderate to good original scores (AB) either reaches a plateau or decreases after just 5 out of 30 available steps. Replotting the sampling trajectories using data driven trial-difficulty estimates (D and DS) indicates that the intended difficulty increments across time were not achieved. Furthermore, the correlation between original scores and accuracy or response time measures also are hard to interpret, suggesting that ceiling effects were not the only problem, as the ordering of the scale may also be sub-optimal for the UK Biobank population, perhaps due to it having been developed with a United States (US) population in mind. This resulted in the original summary scores (AB) having a malformed bimodal distribution with a high mean estimate (0.8 where 1 is maximum), which is unlikely to reflect the true underlying population distribution of crystalised intelligence abilities (25).

IDoTC operated by (i) leveraging the population's accuracy and response time measures while (ii) factoring in the component of performance variance that is better explained by basic visuomotor response times and (iii) taking into account potential (and here intended) bias due to more difficult items being sampled for higher performing abilities. Taken together, these characteristics enable recalculation of the trial-difficulty scale in a data driven manner. The resultant DS scale can potentially be used in future studies as the basis for the adaptive sampling algorithm, though it may still be advisable to include more high-difficulty items given the observed ceiling effects.

More importantly, analysis of the data distributions demonstrates that IDoCT was successful in addressing the issues with the original summary scores. For example, AS has the expected Gaussian distribution (25), supporting the view that IDoCT could overcome the ceiling effect limitation generated by $D_{old}$ and better capture the underlying crystalised language ability that is the target of the task. Plotting the original AB and re-estimated AS scores against the basic mean reaction time and total correct response measures shows distorted distributions for the former but not the latter. Furthermore, by comparing the word difficulty measures before and after the scaling (Figure 5), it is evident that IDoCT properly addresses the biased sampling issue, increasing the measure of difficulty of words that were presented exclusively to participants of higher ability and by decreasing it in the case of words presented only to low performing individuals. Moreover, the retest plots demonstrate a more homogeneous spread on the Bland-Altman plots, with lower SD difference and higher cross-session correlation for AS. In sum, all analyses indicate that the performance score distributions from IDoCT are superior to those output by the original task.

The obtained measure of DT and AS are further validated by the results on the associations with age and education. Specifically, although DT shows the expected increase with age (26), the effect size of the relationship with education level is negligible. This supports the view that DT captures individual differences in fundamental visuo-motor latencies, as opposed to the knowledge of the meaning of words. Conversely, AS improves with age - the expected pattern of results as knowledge of words improves and 'crystalizes' throughout the lifespan (27–30) - but also improves with exposure to higher education, where there is a higher likelihood of learning unusual words (31,32). Notably, the scale of both of these associations is numerically stronger for AS than for the original AB score, together resulting an improved $R^2$ in the linear regression analysis (AB: 0.17, AS: 0.21).

The above results confirm that IDoCT is successful in recalculating the difficulty scale, and in fractioning performance into distinct cognitive and visuomotor components that have superior re-test properties and improved demographic predictive validity. These findings align with our previous applications of this technique to tasks from the Cognitron library (14). The more novel question pertains to whether these advantages in the precision of the task performance estimates extend to improvements in imaging associations.

The machine learning pipeline addresses this by measuring how accurately the original AB measure can be predicted from data of different imaging modalities and using this as a baseline for comparing the AS and DT performance estimates. Overall, the behavioural – imaging associations are in the small range, albeit statistically significant. This may not be unexpected given the simplicity of the linear regression machine learning approach applied and the recent literature on the scale of such associations when estimated within well powered datasets (33). More importantly, the associations with AS are consistently stronger than AB across all analysed imaging modalities. Among the four datasets studied, the volume measures led to models with the highest $R^2$ on the test set both for AS and DT, suggesting that these measures might be more informative when trying to predict cognitive ability and visuo-motor latency.

The NLP analysis provides an unbiased data driven way to qualitatively evaluate the functional specificity of the AS and DT measures, by determining the most common functional terms that their associated brain regions co-occur with within literature. The results have face value, with brain regions identified as best predictors of DT mainly relating to visual and motor functions, such as the Intracalcarine cortex (34), and the cerebellum (35). This is expected considered that DT is supposed to measure visuo-motor latency times. Conversely, AS is mainly associated with brain regions involved in memory, language and auditory functions, such as the hippocampus (36), uncinate fasciculus (37) and inferior temporal gyrus (38). Considering the nature of the task, which requires participants to associate the meaning of spoken/written words to different images, the identified brain functions are as expected.

A strength of this study is the sample size, which allows to draw firmer and more precise conclusions. An important aspect to consider, however, is that the UK Biobank sample includes mainly middle to older aged individuals (<1% below 50 years old), that are not necessarily representative of the general population in terms of health, physical, and lifestyle aspects (39). This does not undermine our validation of the IDoCT approach, but it should be noted that the data-driven measures of trials' difficulty might change if younger individuals are included in the analysis, as different words are more likely to be learnt at different stages of the lifespan (e.g., school vs work). A further strength of the study is that four different kinds of features (imaging-derived phenotypes) from two imaging data modalities were analysed, in combination with both behavioural and demographics data, which allows to gain better insight into the best imaging predictors of AS and DT.

Despite these strengths, there are some limitations. First, IDoCT requires as input a trial's description in order to extract data-driven measures of trials' difficulty. In this study, the word presented at each trial is used as that definition because it was the only information available. However, in the case of PVT, the difficulty of a trial is influenced not only by the word presented, but also by the set of pictures from which participants are required to choose. Without this information, it is difficult to interpret the reason why specific words were assigned higher or lower difficulty scores, as the motivation could be a combination of the difficulty of the word itself or of the figures presented. However, a general observation is that higher values of DS on the difficulty scale are mainly characterised by a Latin, French or Greek etymology. This pattern of results again has some face validity, because English is not a romance, but a Germanic language (40), but corresponds only to a limited interpretation due to the lack of information about the word-figure pairs.

Further limitations are related to the NLP analysis. First, the latter was conducted exclusively on open access papers, which limits the literature research to previous studies that are freely accessible. In addition, although the NLP analysis conducted in this study was sufficient to achieve the intended aim, it consisted exclusively of the estimation of word frequency across different papers. Further research could extend this by using more advanced NLP approaches, such as topic modelling to extract common topics (41), or by implementing digraphs and dependency parsing to identify related and connected words (42). The latter approach could provide additional information on the meaning of the words within the context of the sentence in which they appear and help in reducing the noise of the one to many mappings between brain structures and cognitive functions.

Finally, there are two limitations of the imaging analysis that could be addressed by further research. First, the study is exclusively focused on of associations with structural measures. However, cognitive processes generally result from complex brain networks (43), rather than individual discrete contributions of brain regions (44). Analysis using graph theoretic approaches to quantifying the information processing properties of networks, or focused on network dynamics from functional MRI might provide better and more detailed insights into the neural correlates of cognitive task performance. Relatedly, the machine learning approaches used here are mainly state-of-the-art shallow linear models and they are applied independently to each imaging modality. The strength of associations could improve if more advanced methods, such as deep learning and combination of multi-modal features are applied. Nonetheless, the fact that association strengths increase for AS vs AB across all imaging modalities confirms our primary hypothesis that IDoCT can provide superior cognitive ability measures for future association studies, including those investigating more advanced imaging analysis methods.

In conclusion, we successfully apply IDoCT to the PVT data collected as part of the UK Biobank imaging extension, and obtain superior prediction of subject-level cognitive ability and visuo-motor latency, as well as an optimised data-driven word difficulty scale calibrated on the UK population. Our results further validate IDoCT by showing the improved relationship between the performance metrics and age and education, as well as brain imaging metrics in terms of the strength and functional specificity of associations.

## 5. ACKNOWLEDGMENTS

V.G. was supported by the NIHR Imperial Biomedical Research Centre (BRC) grant to AH and by the Medical Research Council, MR/W00710X/1. NA was funded by the Medical Research Council and Wellcome Trust.

## 6. COMPETING INTERESTS

The authors declare no competing interests.

# APPENDIX A

**Table A1 List of structural MRI (sMRI) imaging features**. The imaging data consisted of 26 measures of fractional anisotropy (FA) from diffusion weighted images (DWI), and 139 volume, 43 intensity and 60 cortical thickness measures from T1-weighted structural magnetic resonance images (MRI). The imaging features included in the study are provided in the Table.

| Dataset | Features |
|---|---|
| **FA** | • Tract anterior thalamic radiation (right)<br>• Tract cingulate gyrus part of cingulum (right)<br>• Tract corticospinal tract (right)<br>• Tract inferior longitudinal fasciculus (right)<br>• Tract medial lemniscus (right)<br>• Tract posterior thalamic radiation (right)<br>• Tract superior thalamic radiation (right)<br>• Tract anterior thalamic radiation (left)<br>• Tract cingulate gyrus part of cingulum (left)<br>• Tract corticospinal tract (left)<br>• Tract inferior longitudinal fasciculus (left)<br>• Tract posterior thalamic radiation (left)<br>• Tract superior longitudinal fasciculus (left)<br>• Tract uncinate fasciculus (left)<br>• Tract acoustic radiation (left)<br>• Tract parahippocampal part of cingulum (left)<br>• Tract forceps major<br>• Tract inferior fronto occipital fasciculus (left)<br>• Tract medial lemniscus (left)<br>• Tract superior thalamic radiation (left)<br>• Tract acoustic radiation (right)<br>• Tract parahippocampal part of cingulum (right)<br>• Tract forceps minor<br>• Tract inferior fronto occipital fasciculus (right)<br>• Tract middle cerebellar peduncle<br>• Tract superior longitudinal fasciculus (right) |
| **Thickness** | • Cuneus (left hemisphere)<br>• Lateraloccipital (left hemisphere)<br>• Lingual (left hemisphere)<br>• Parsorbitalis (left hemisphere)<br>• Posteriorcingulate (left hemisphere)<br>• Precuneus (left hemisphere)<br>• Superiorparietal (left hemisphere)<br>• Entorhinal (right hemisphere)<br>• Isthmuscingulate (right hemisphere)<br>• Lateralorbitofrontal (right hemisphere)<br>• Parsopercularis (right hemisphere)<br>• Parstriangularis (right hemisphere)<br>• Precentral (right hemisphere)<br>• Superiortemporal (right hemisphere)<br>• Entorhinal (left hemisphere)<br>• Lateralorbitofrontal (left hemisphere)<br>• Medialorbitofrontal (left hemisphere)<br>• Parstriangularis (left hemisphere)<br>• Precentral (left hemisphere)<br>• Rostralanteriorcingulate (left hemisphere)<br>• Superiortemporal (left hemisphere)<br>• Fusiform (right hemisphere)<br>• Lateraloccipital (right hemisphere)<br>• Lingual (right hemisphere)<br>• Parsorbitalis (right hemisphere)<br>• Pericalcarine (right hemisphere)<br>• Precuneus (right hemisphere) |
| | • Supramarginal (right hemisphere)<br>• Caudalmiddlefrontal (left hemisphere)<br>• Inferiorparietal (left hemisphere)<br>• Isthmuscingulate (left hemisphere)<br>• Parahippocampal (left hemisphere)<br>• Parsopercularis (left hemisphere)<br>• Postcentral (left hemisphere)<br>• Superiorfrontal (left hemisphere)<br>• Transversetemporal (left hemisphere)<br>• Caudalanteriorcingulate (right hemisphere)<br>• Cuneus (right hemisphere)<br>• Inferiortemporal (right hemisphere)<br>• Middletemporal (right hemisphere)<br>• Paracentral (right hemisphere)<br>• Posteriorcingulate (right hemisphere)<br>• Rostralmiddlefrontal (right hemisphere)<br>• Superiorparietal (right hemisphere)<br>• Transversetemporal (right hemisphere)<br>• Caudalanteriorcingulate (left hemisphere)<br>• Fusiform (left hemisphere)<br>• Inferiortemporal (left hemisphere)<br>• Middletemporal (left hemisphere)<br>• Paracentral (left hemisphere)<br>• Pericalcarine (left hemisphere)<br>• Rostralmiddlefrontal (left hemisphere)<br>• Supramarginal (left hemisphere)<br>• Caudalmiddlefrontal (right hemisphere)<br>• Inferiorparietal (right hemisphere)<br>• Medialorbitofrontal (right hemisphere)<br>• Parahippocampal (right hemisphere)<br>• Postcentral (right hemisphere)<br>• Rostralanteriorcingulate (right hemisphere)<br>• Superiorfrontal (right hemisphere) |
| **Volume** | • Insular Cortex (left)<br>• Inferior Frontal Gyrus pars triangularis (left)<br>• Precentral Gyrus (left)<br>• Temporal Pole (left)<br>• Superior Temporal Gyrus posterior division (left)<br>• Middle Temporal Gyrus temporooccipital part (left)<br>• Inferior Temporal Gyrus anterior division (left)<br>• Postcentral Gyrus (left)<br>• Supramarginal Gyrus anterior division (left)<br>• Supramarginal Gyrus posterior division (left)<br>• Lateral Occipital Cortex superior division (right)<br>• Frontal Medial Cortex (right)<br>• Subcallosal Cortex (right)<br>• Paracingulate Gyrus (right)<br>• Precuneous Cortex (right)<br>• Cuneal Cortex (right)<br>• Parahippocampal Gyrus anterior division (right)<br>• Temporal Fusiform Cortex anterior division (right)<br>• Temporal Fusiform Cortex posterior division (left)<br>• Frontal Operculum Cortex (left)<br>• Parietal Operculum Cortex (left)<br>• Planum Polare (left)<br>• Supracalcarine Cortex (left)<br>• Occipital Pole (left)<br>• Caudate (left)<br>• Hippocampus (left)<br>• Ventral Striatum (right)<br>• I IV Cerebellum (left) |
| | • VI Cerebellum (left)<br>• Crus II Cerebellum (left)<br>• Crus II Cerebellum (right)<br>• VIIIa Cerebellum (right)<br>• IX Cerebellum (left)<br>• IX Cerebellum (right)<br>• Insular Cortex (right)<br>• Inferior Frontal Gyrus pars triangularis (right)<br>• Precentral Gyrus (right)<br>• Temporal Pole (right)<br>• Superior Temporal Gyrus posterior division (right)<br>• Middle Temporal Gyrus temporooccipital part (right)<br>• Inferior Temporal Gyrus anterior division (right)<br>• Postcentral Gyrus (right)<br>• Supramarginal Gyrus anterior division (right)<br>• Supramarginal Gyrus posterior division (right)<br>• Lateral Occipital Cortex superior division (left)<br>• Frontal Medial Cortex (left)<br>• Subcallosal Cortex (left)<br>• Paracingulate Gyrus (left)<br>• Precuneous Cortex (left)<br>• Cuneal Cortex (left)<br>• Parahippocampal Gyrus anterior division (left)<br>• Temporal Fusiform Cortex anterior division (left)<br>• Temporal Fusiform Cortex posterior division (right)<br>• Frontal Operculum Cortex (right)<br>• Parietal Operculum Cortex (right)<br>• Planum Polare (right)<br>• Supracalcarine Cortex (right)<br>• Occipital Pole (right)<br>• Caudate (right)<br>• Hippocampus (right)<br>• Ventral Striatum (left)<br>• Brain Stem<br>• V Cerebellum (right)<br>• Crus I Cerebellum (right)<br>• Crus II Cerebellum (vermis)<br>• VIIIa Cerebellum (vermis)<br>• VIIIb Cerebellum (right)<br>• IX Cerebellum (vermis)<br>• X Cerebellum (right)<br>• Frontal Pole (right)<br>• Superior Frontal Gyrus (right)<br>• Middle Frontal Gyrus (right)<br>• Inferior Frontal Gyrus pars opercularis (right)<br>• Superior Temporal Gyrus anterior division (right)<br>• Middle Temporal Gyrus anterior division (right)<br>• Middle Temporal Gyrus posterior division (right)<br>• Inferior Temporal Gyrus posterior division (right)<br>• Inferior Temporal Gyrus temporooccipital part (right)<br>• Superior Parietal Lobule (right)<br>• Angular Gyrus (right)<br>• Lateral Occipital Cortex inferior division (left)<br>• Intracalcarine Cortex (left)<br>• Juxtapositional Lobule Cortex formerly Supplementary Motor Cortex (left)<br>• Cingulate Gyrus anterior division (left)<br>• Cingulate Gyrus posterior division (left)<br>• Frontal Orbital Cortex (left)<br>• Parahippocampal Gyrus posterior division (left)<br>• Lingual Gyrus (left)<br>• Temporal Occipital Fusiform Cortex (right) |
| | • Occipital Fusiform Gyrus (right)<br>• Central Opercular Cortex (right)<br>• Heschl s Gyrus includes H1 and H2 (right)<br>• Planum Temporale (right)<br>• Thalamus (right)<br>• Putamen (right)<br>• Pallidum (right)<br>• Amygdala (left)<br>• I IV Cerebellum (right)<br>• VI Cerebellum (vermis)<br>• Crus I Cerebellum (left)<br>• VIIb Cerebellum (left)<br>• VIIb Cerebellum (right)<br>• VIIIb Cerebellum (left)<br>• X Cerebellum (left)<br>• Frontal Pole (left)<br>• Superior Frontal Gyrus (left)<br>• Middle Frontal Gyrus (left)<br>• Inferior Frontal Gyrus pars opercularis (left)<br>• Superior Temporal Gyrus anterior division (left)<br>• Middle Temporal Gyrus anterior division (left)<br>• Middle Temporal Gyrus posterior division (left)<br>• Inferior Temporal Gyrus posterior division (left)<br>• Inferior Temporal Gyrus temporooccipital part (left)<br>• Superior Parietal Lobule (left)<br>• Angular Gyrus (left)<br>• Lateral Occipital Cortex inferior division (right)<br>• Intracalcarine Cortex (right)<br>• Juxtapositional Lobule Cortex formerly Supplementary Motor Cortex (right)<br>• Cingulate Gyrus anterior division (right)<br>• Cingulate Gyrus posterior division (right)<br>• Frontal Orbital Cortex (right)<br>• Parahippocampal Gyrus posterior division (right)<br>• Lingual Gyrus (right)<br>• Temporal Occipital Fusiform Cortex (left)<br>• Occipital Fusiform Gyrus (left)<br>• Central Opercular Cortex (left)<br>• Heschl s Gyrus includes H1 and H2 (left)<br>• Planum Temporale (left)<br>• Thalamus (left)<br>• Putamen (left)<br>• Pallidum (left)<br>• Amygdala (right)<br>• V Cerebellum (left)<br>• VI Cerebellum (right)<br>• Crus I Cerebellum (vermis)<br>• VIIb Cerebellum (vermis)<br>• VIIIa Cerebellum (left)<br>• VIIIb Cerebellum (vermis)<br>• X Cerebellum (vermis) |
| **Intensity** | • Brain Stem (whole brain)<br>• CC Mid Posterior (whole brain)<br>• CC Mid Anterior (whole brain)<br>• Inf Lat Vent (left hemisphere)<br>• Cerebellum Cortex (left hemisphere)<br>• Amygdala (left hemisphere)<br>• VentralDC (left hemisphere)<br>• Lateral Ventricle (right hemisphere)<br>• Putamen (right hemisphere)<br>• Hippocampus (right hemisphere)<br>• 3rd Ventricle (whole brain) |
| | • CSF (whole brain)<br>• CC Central (whole brain)<br>• CC Anterior (whole brain)<br>• Lateral Ventricle (left hemisphere)<br>• Cerebellum White Matter (left hemisphere)<br>• Hippocampus (left hemisphere)<br>• Accumbens area (left hemisphere)<br>• Caudate (right hemisphere)<br>• Pallidum (right hemisphere)<br>• Choroid plexus (right hemisphere)<br>• 5th Ventricle (whole brain)<br>• Non WM hypointensities (whole brain)<br>• CC Posterior (whole brain)<br>• Thalamus Proper (left hemisphere)<br>• Putamen (left hemisphere)<br>• Vessel (left hemisphere)<br>• Inf Lat Vent (right hemisphere)<br>• Cerebellum Cortex (right hemisphere)<br>• Amygdala (right hemisphere)<br>• VentralDC (right hemisphere)<br>• 4th Ventricle (whole brain)<br>• WM hypointensities (whole brain)<br>• Optic Chiasm (whole brain)<br>• Caudate (left hemisphere)<br>• Pallidum (left hemisphere)<br>• Choroid plexus (left hemisphere)<br>• Cerebellum White Matter (right hemisphere)<br>• Thalamus Proper (right hemisphere)<br>• Accumbens area (right hemisphere)<br>• Vessel (right hemisphere) |

**Table A2 PubMed Central search criteria**. Different search criteria were used to identify previous literature associated to the brain labels of the neural correlates of AS an DT. The list of the search criteria, together with the number of the papers that matched those criteria, is provided in the Table.

| Brain region | PubMed search criteria | Number of papers |
|---|---|---|
| Hippocampus | (hippocampus[Title]) AND function[Title] | 116 |
| Superior temporal gyrus | ((superior temporal gyrus[Title]) AND function) | 85 |
| Transverse temporal gyrus | (transverse temporal gyrus[Title]) OR ((Heschl's gyrus[Title]) AND function) | 25 |
| Medial Lemniscus | ((((medial lemniscus[Title]) OR Reil's ribbon[Title]) OR Reil's band[Title])) OR medial lemniscus[Abstract] | 99 |
| Anterior corpus callosum | ((corpus callosum[Title]) AND anterior[Abstract]) AND cognitive function | 114 |
| Medial Frontal Cortex | medial frontal cortex[Title] | 102 |
| Uncinate fasciculus | ((uncinate fasciculus[Abstract]) AND uncinate fasciculus[Title]) AND function | 47 |
| Nucleus Accumbens | (accumbens[Title]) AND cognition[Body - Key Terms] | 89 |
| Caudal middle frontal gyrus | (((caudal middle frontal[Title]) OR caudal middle frontal[Abstract])) AND function | 65 |
| Inferior temporal gyrus | inferior temporal gyrus[Title] | 12 |
| Superior frontal gyrus | ((superior frontal gyrus[Title]) AND function) | 22 |
| Parahippocampal part of cingulum | (((parahippocampal cingulum[Abstract]) OR parahippocampal cingulum[Title])) AND function | 38 |
| Forceps major | (((forceps major[Abstract]) OR forceps major[Title])) AND forceps major[Body - Key Terms] | 92 |
| Posterior thalamic radiations | (((posterior thalamic radiation[Abstract]) OR posterior thalamic radiation[Title])) AND function | 105 |
| Middle temporal gyrus | middle temporal area[Title] | 31 |
| Lateral orbitofrontal cortex | lateral orbitofrontal cortex[Title] | 37 |
| Occipital fusiform gyrus | (occipital fusiform gyrus[Abstract]) OR occipital fusiform gyrus[Title] | 19 |
| Lateral occipital cortex | (lateral occipital cortex[Title]) AND function | 19 |
| Amygdala | (amygdala[Title]) AND function[Title] | 85 |
| Frontal pole/lobe | ((frontal lobe[Title]) AND cognition[Body - Key Terms]) | 97 |
| Insular cortex | (((insular cortex[Title]))) AND cognition[Body - All Words] | 62 |
| Cerebellum | ((cerebellum[Title])) AND cognitive function[Body - Key Terms] | 117 |
| Temporal fusiform cortex | (((temporal fusiform cortex[Title]) AND cognition[Body - Key Terms])) OR temporal fusiform cortex[Abstract] | 12 |
| Intra calcarine cortex | ((intra-calcarine[Abstract]) OR intracalcarine[Abstract]) AND function[Body - All Words] | 19 |
| Occipital pole/lobe | occipital lobe[Title] | 93 |

**Table A3 IDoCT results of D and DS**. List of words used in the PVT Task and the measures of D and DS derived from IDoCT.

| Meaning | D | DS | Meaning | D | DS | Meaning | D | DS | Meaning | D | DS |
|---|---|---|---|---|---|---|---|---|---|---|---|
| **crayon** | 0,52 | 0,52 | **confine** | 0,57 | 0,74 | **shore** | 0,51 | 0,7 | **adhere** | 0,67 | 0,69 |
| **nap** | 0,52 | 0,52 | **corroded** | 0,53 | 0,72 | **shovel** | 0,55 | 0,74 | **amiable** | 0,59 | 0,61 |
| **paper** | 0,6 | 0,61 | **equation** | 0,54 | 0,73 | **tent** | 0,59 | 0,6 | **arid** | 0,6 | 0,65 |
| **party** | 0,61 | 0,71 | **erode** | 0,59 | 0,65 | **belt** | 0,6 | 0,65 | **bestow** | 0,67 | 0,68 |
| **sick** | 0,59 | 0,61 | **fabricate** | 0,58 | 0,51 | **braid** | 0,57 | 0,75 | **cascade** | 0,54 | 0,71 |
| **big** | 0,58 | 0,63 | **frothy** | 0,58 | 0,61 | **broom** | 0,53 | 0,74 | **chaff** | 0,62 | 0,65 |
| **bus** | 0,53 | 0,71 | **gavel** | 0,62 | 0,76 | **brush** | 0,52 | 0,73 | **concave** | 0,68 | 0,66 |
| **egg** | 0,65 | 0,63 | **journal** | 0,62 | 0,82 | **buy** | 0,54 | 0,73 | **disarray** | 0,56 | 0,75 |
| **fun** | 0,61 | 0,62 | **knead** | 0,54 | 0,66 | **cactus** | 0,53 | 0,74 | **excavate** | 0,56 | 0,69 |
| **orange** | 0,66 | 0,73 | **ladle** | 0,52 | 0,67 | **hive** | 0,52 | 0,74 | **irate** | 0,54 | 0,64 |
| **tree** | 0,62 | 0,79 | **manger** | 0,76 | 0,75 | **lighthouse** | 0,51 | 0,69 | **jocular** | 0,56 | 0,68 |
| **triangle** | 0,53 | 0,72 | **masonry** | 0,59 | 0,64 | **mask** | 0,7 | 0,67 | **lustre** | 0,64 | 0,66 |
| **tricycle** | 0,52 | 0,71 | **molten** | 0,54 | 0,73 | **melted** | 0,56 | 0,77 | **offspring** | 0,57 | 0,77 |
| **under** | 0,54 | 0,69 | **odometer** | 0,73 | 0,74 | **railroad** | 0,57 | 0,77 | **ravine** | 0,65 | 0,69 |
| **zoo** | 0,56 | 0,7 | **oxidized** | 0,59 | 0,64 | **rubbish** | 0,59 | 0,63 | **receptacle** | 0,63 | 0,67 |
| **fireman** | 0,63 | 0,71 | **quibble** | 0,57 | 0,71 | **blanket** | 0,53 | 0,73 | **refurbish** | 0,55 | 0,61 |
| **jar** | 0,52 | 0,74 | **rural** | 0,61 | 0,65 | **bucket** | 0,56 | 0,71 | **refuse** | 0,68 | 0,64 |
| **log** | 0,51 | 0,72 | **satchel** | 0,52 | 0,74 | **calm** | 0,52 | 0,72 | **summit** | 0,58 | 0,6 |
| **pan** | 0,51 | 0,73 | **scholar** | 0,61 | 0,73 | **carry** | 0,53 | 0,73 | **abode** | 0,59 | 0,65 |
| **queen** | 0,52 | 0,75 | **sentry** | 0,57 | 0,7 | **castle** | 0,52 | 0,75 | **cubicle** | 0,63 | 0,76 |
| **run** | 0,58 | 0,59 | **shimmer** | 0,58 | 0,74 | **cattle** | 0,55 | 0,74 | **decrepit** | 0,54 | 0,73 |
| **ride** | 0,69 | 0,77 | **tarpaulin** | 0,55 | 0,62 | **fancy** | 0,63 | 0,77 | **effervescent** | 0,62 | 0,63 |
| **rocket** | 0,51 | 0,73 | **trench** | 0,52 | 0,74 | **metal** | 0,51 | 0,73 | **fiery** | 0,61 | 0,65 |
| **stick** | 0,64 | 0,63 | **unstable** | 0,54 | 0,72 | **net** | 0,52 | 0,73 | **figurine** | 0,53 | 0,73 |
| **tractor** | 0,53 | 0,74 | **attire** | 0,6 | 0,65 | **plastic** | 0,52 | 0,65 | **gregarious** | 0,57 | 0,62 |
| **wet** | 0,64 | 0,66 | **barren** | 0,59 | 0,7 | **porch** | 0,52 | 0,72 | **hydroponics** | 0,69 | 0,68 |
| **airplane** | 0,53 | 0,66 | **beacon** | 0,59 | 0,62 | **port** | 0,51 | 0,71 | **jovial** | 0,55 | 0,63 |
| **angry** | 0,65 | 0,56 | **broth** | 0,64 | 0,84 | **acrobat** | 0,53 | 0,72 | **labryinth** | 0,63 | 0,76 |
| **back (of)** | 0,54 | 0,73 | **camouflaged** | 0,55 | 0,78 | **cottage** | 0,51 | 0,69 | **noxious** | 0,56 | 0,74 |
| **beside** | 0,58 | 0,79 | **comrade** | 0,73 | 0,74 | **diamond** | 0,61 | 0,8 | **pachyderm** | 0,87 | 0,77 |
| **chair** | 0,68 | 0,75 | **frigid** | 0,67 | 0,75 | **envelope** | 0,51 | 0,72 | **panorama** | 0,52 | 0,65 |
| **coat** | 0,58 | 0,63 | **knick-knack** | 0,62 | 0,78 | **fellowship** | 0,52 | 0,67 | **quills** | 0,52 | 0,69 |
| **doctor** | 0,56 | 0,73 | **pageantry** | 0,69 | 0,76 | **gold** | 0,52 | 0,73 | **resplendent** | 0,58 | 0,63 |
| **fly** | 0,57 | 0,59 | **tarnished** | 0,62 | 0,73 | **instrument** | 0,52 | 0,74 | **silhouette** | 0,63 | 0,68 |
| **leaf** | 0,69 | 0,76 | **tattered** | 0,56 | 0,75 | **jewelry** | 0,51 | 0,72 | **slat** | 0,57 | 0,64 |
| **ocean** | 0,57 | 0,66 | **thicket** | 0,61 | 0,65 | **monument** | 0,57 | 0,71 | **striated** | 0,7 | 0,7 |
| **open** | 0,54 | 0,75 | **transact** | 0,56 | 0,7 | **pyramid** | 0,51 | 0,69 | **vortex** | 0,55 | 0,75 |
| **palace** | 0,52 | 0,73 | **turbulent** | 0,59 | 0,67 | **rose** | 0,54 | 0,74 | **celestial** | 0,64 | 0,66 |
| **riding** | 0,52 | 0,74 | **tusk** | 0,51 | 0,73 | **tortoise** | 0,51 | 0,69 | **cosmopolitan** | 0,63 | 0,66 |
| **voyage** | 0,52 | 0,74 | **depot** | 0,55 | 0,74 | **construct** | 0,54 | 0,76 | **trivet** | 0,76 | 0,87 |
| **windmill** | 0,52 | 0,75 | **distinguished** | 0,6 | 0,7 | **creek** | 0,62 | 0,83 | **verdant** | 0,65 | 0,68 |
| **garden** | 0,52 | 0,71 | **flamboyant** | 0,57 | 0,72 | **ferocious** | 0,52 | 0,7 | **affable** | 0,62 | 0,69 |
| **boulder** | 0,55 | 0,74 | **foal** | 0,56 | 0,62 | **flimsy** | 0,54 | 0,72 | **cherubic** | 0,63 | 0,64 |
| **bouquet** | 0,51 | 0,68 | **lounging** | 0,53 | 0,69 | **hedge** | 0,53 | 0,74 | **coalesce** | 0,74 | 0,73 |
| **cable** | 0,53 | 0,69 | **tethered** | 0,55 | 0,73 | **incinerate** | 0,58 | 0,74 | **deluge** | 0,59 | 0,63 |
| **exhibit** | 0,6 | 0,78 | **zenith** | 0,74 | 0,73 | **instruct** | 0,53 | 0,75 | **didactic** | 0,73 | 0,84 |
| **haul** | 0,59 | 0,69 | **baubles** | 0,58 | 0,73 | **lobby** | 0,59 | 0,76 | **embellish** | 0,74 | 0,67 |
| **hilarious** | 0,57 | 0,68 | **buffoon** | 0,71 | 0,71 | **mine** | 0,57 | 0,74 | **encumber** | 0,69 | 0,72 |
| **indicate** | 0,55 | 0,69 | **burly** | 0,58 | 0,65 | **orchard** | 0,52 | 0,72 | **fissure** | 0,62 | 0,69 |
| **launch** | 0,57 | 0,75 | **cleave** | 0,6 | 0,67 | **peak** | 0,58 | 0,66 | **hodgepodge** | 0,78 | 0,73 |
| **marsh** | 0,59 | 0,68 | **domicile** | 0,64 | 0,68 | **prohibit** | 0,54 | 0,7 | **lassitude** | 0,63 | 0,77 |
| **mend** | 0,61 | 0,75 | **horizontal** | 0,74 | 0,72 | **scorched** | 0,82 | 0,79 | **malefactor** | 0,88 | 0,79 |
| **quarrel** | 0,55 | 0,6 | **jowl** | 0,6 | 0,63 | **shatter** | 0,52 | 0,73 | **matron** | 0,86 | 0,81 |
| **royal** | 0,52 | 0,74 | **knoll** | 0,65 | 0,69 | **transparent** | 0,64 | 0,71 | **melancholy** | 0,69 | 0,69 |
| **souvenir** | 0,56 | 0,73 | **ledger** | 0,59 | 0,65 | **trinket** | 0,81 | 0,73 | **onerous** | 0,57 | 0,72 |
| **stubborn** | 0,54 | 0,76 | **lethargic** | 0,61 | 0,66 | **tutor** | 0,52 | 0,74 | **prone** | 0,7 | 0,71 |
| **absorb** | 0,58 | 0,74 | **orifice** | 0,65 | 0,67 | **utensil** | 0,53 | 0,75 | **revel** | 0,63 | 0,64 |
| **antenna** | 0,52 | 0,74 | **pennant** | 0,68 | 0,72 | **boulevard** | 0,54 | 0,74 | **scintillating** | 0,67 | 0,68 |
| **blueprint** | 0,6 | 0,72 | **precipice** | 0,67 | 0,66 | **debris** | 0,59 | 0,71 | **surly** | 0,63 | 0,66 |
| **carriage** | 0,53 | 0,71 | **spry** | 0,72 | 0,73 | **discouraged** | 0,56 | 0,73 | **atoll** | 0,57 | 0,63 |
| **citrus** | 0,55 | 0,7 | **wallowing** | 0,59 | 0,67 | **drought** | 0,54 | 0,72 | **cataract** | 0,67 | 0,81 |
| **diagram** | 0,55 | 0,76 | **alms** | 0,7 | 0,7 | **fragment** | 0,55 | 0,7 | **disparate** | 0,65 | 0,79 |
| **dissolved** | 0,53 | 0,75 | **balustrade** | 0,59 | 0,65 | **hurl** | 0,55 | 0,6 | **ebullience** | 0,56 | 0,72 |
| **elastic** | 0,55 | 0,71 | **ellipse** | 0,74 | 0,74 | **judiciary** | 0,53 | 0,71 | **flaccid** | 0,67 | 0,71 |
| **engraved** | 0,52 | 0,74 | **glower** | 0,92 | 0,76 | **morsel** | 0,72 | 0,69 | **hirsute** | 0,56 | 0,74 |
| **festive** | 0,53 | 0,71 | **helix** | 0,73 | 0,7 | **pillars** | 0,54 | 0,74 | **hovel** | 0,56 | 0,6 |
| **herd** | 0,52 | 0,74 | **indisposed** | 0,63 | 0,64 | **reap** | 0,57 | 0,62 | **inclement** | 0,62 | 0,7 |
| **crossroads** | 0,52 | 0,72 | **iridescent** | 0,77 | 0,77 | **residential** | 0,59 | 0,78 | **infuse** | 0,55 | 0,6 |
| **lotion** | 0,52 | 0,74 | **mirth** | 0,62 | 0,67 | **slit** | 0,62 | 0,69 | **obstreperous** | 0,72 | 0,72 |
| **luggage** | 0,51 | 0,7 | **opulent** | 0,59 | 0,64 | **snarl** | 0,54 | 0,72 | **paucity** | 0,61 | 0,73 |
| **pamphlet** | 0,58 | 0,73 | **parry** | 0,66 | 0,68 | **stampede** | 0,52 | 0,72 | **pecuniary** | 0,59 | 0,74 |
| **quartet** | 0,54 | 0,72 | **perusal** | 0,66 | 0,74 | **startled** | 0,52 | 0,66 | **plethora** | 0,85 | 0,75 |
| **ripple** | 0,51 | 0,73 | **quadruped** | 0,82 | 0,81 | **submerged** | 0,61 | 0,74 | **potable** | 0,65 | 0,79 |
| **sapling** | 0,57 | 0,66 | **quagmire** | 0,57 | 0,61 | **weld** | 0,52 | 0,65 | **prodigious** | 0,74 | 0,87 |
| **sculpture** | 0,51 | 0,71 | **recess** | 0,69 | 0,7 | **barricade** | 0,54 | 0,69 | **progeny** | 0,67 | 0,77 |
| **transport** | 0,57 | 0,76 | **ruminate** | 0,59 | 0,65 | **bureau** | 0,73 | 0,7 | **repast** | 0,58 | 0,72 |
| **trophy** | 0,6 | 0,75 | **shorn** | 0,64 | 0,67 | **congestion** | 0,53 | 0,61 | **sinuous** | 0,69 | 0,71 |
| **vineyard** | 0,56 | 0,63 | **trend** | 0,66 | 0,71 | **consume** | 0,58 | 0,68 | **sumptuous** | 0,62 | 0,68 |
| **cove** | 0,56 | 0,7 | **truncate** | 0,74 | 0,85 | hangar | 0,52 | 0,74 | **detritus** | 0,55 | 0,7 |
| **descend** | 0,61 | 0,62 | **abrade** | 0,68 | 0,8 | memento | 0,7 | 0,7 | **diadem** | 0,7 | 0,82 |
| **desolate** | 0,52 | 0,65 | **anomalous** | 0,78 | 0,74 | monarch | 0,53 | 0,58 | **egress** | 0,6 | 0,75 |
| **dune** | 0,56 | 0,72 | **bucolic** | 0,76 | 0,87 | shabby | 0,53 | 0,73 | **feral** | 0,66 | 0,66 |
| **ensemble** | 0,79 | 0,76 | **buffet** | 0,79 | 0,92 | tranquil | 0,56 | 0,64 | **sustenance** | 0,71 | 0,69 |
| **histrionic** | 0,61 | 0,75 | **delineation** | 0,7 | 0,71 | acoustic | 0,7 | 0,66 | **tractable** | 0,69 | 0,8 |
| **natty** | 0,56 | 0,72 | **fecund** | 0,67 | 0,82 | brawny | 0,6 | 0,74 | **concomitant** | 0,74 | 0,72 |

**Table A4 Feature selection of FA dataset when predicting AS**. (P-2) linear regression models were trained (with P equal to the number of features in the dataset) using 5-fold cross validation and at each iteration the feature with the lowest correlation with AS was removed from the regressors. The mean $R^2$ in the train and test set of each model across the folds, as well as the name of the feature dropped at each iteration, is reported in the Table.

| **Model Number** | **Features dropped** | **AS** | |
|---|---|---|---|
| | | **Test R2 (mean ± std)** | **Train R2 (mean ± std)** |
| **c** | **FA in tract cingulate gyrus part of cingulum right** | **0.009±0.003** | **0.012±0.001** |
| 1 | FA in tract acoustic radiation right | 0.009±0.003 | 0.012±0.001 |
| 2 | FA in tract forceps major | 0.009±0.003 | 0.011±0.001 |
| 3 | FA in tract middle cerebellar peduncle | 0.009±0.003 | 0.011±0.001 |
| 4 | FA in tract cingulate gyrus part of cingulum left | 0.009±0.003 | 0.011±0.001 |
| 5 | FA in tract acoustic radiation left | 0.009±0.003 | 0.011±0.001 |
| 6 | FA in tract inferior fronto occipital fasciculus left | 0.009±0.003 | 0.011±0.001 |
| 7 | FA in tract forceps minor | 0.008±0.003 | 0.011±0.001 |
| 8 | FA in tract inferior longitudinal fasciculus left | 0.008±0.003 | 0.01±0.001 |
| 9 | FA in tract inferior fronto occipital fasciculus right | 0.008±0.003 | 0.01±0.001 |
| 10 | FA in tract parahippocampal part of cingulum left | 0.008±0.003 | 0.009±0.001 |
| 11 | FA in tract inferior longitudinal fasciculus right | 0.008±0.003 | 0.009±0.001 |
| 12 | FA in tract superior longitudinal fasciculus left | 0.007±0.004 | 0.009±0.001 |
| 13 | FA in tract posterior thalamic radiation left | 0.007±0.003 | 0.009±0.001 |
| 14 | FA in tract superior longitudinal fasciculus right | 0.007±0.003 | 0.009±0.001 |
| 15 | FA in tract corticospinal tract left | 0.007±0.003 | 0.009±0.001 |
| 16 | FA in tract parahippocampal part of cingulum right | 0.007±0.003 | 0.008±0.001 |
| 17 | FA in tract anterior thalamic radiation right | 0.007±0.003 | 0.008±0.001 |
| 18 | FA in tract posterior thalamic radiation right | 0.007±0.003 | 0.008±0.001 |
| 19 | FA in tract corticospinal tract right | 0.007±0.003 | 0.008±0.001 |
| 20 | FA in tract uncinate fasciculus left | 0.007±0.003 | 0.008±0.001 |
| 21 | FA in tract anterior thalamic radiation left | 0.007±0.003 | 0.007±0.001 |
| 22 | FA in tract medial lemniscus left | 0.006±0.003 | 0.007±0.001 |
| 23 | FA in tract superior thalamic radiation right | 0.006±0.003 | 0.006±0.001 |
| 24 | FA in tract superior thalamic radiation left | 0.006±0.002 | 0.006±0.0 |

**Table A5 Feature selection of thickness dataset when predicting AS**. (P-2) linear regression models were trained (with P equal to the number of features in the dataset) using 5-fold cross validation and at each iteration the feature with the lowest correlation with AS was removed from the regressors. The mean $R^2$ in the train and test set of each model across the folds, as well as the name of the feature dropped at each iteration, is reported in the Table.

| **Model Number** | **Features dropped** | **AS** | |
|---|---|---|---|
| | | **Test R2 (mean ± std)** | **Train R2 (mean ± std)** |
| **0** | **Mean thickness of parstriangularis right hemisphere** | **0.01±0.008** | **0.017±0.002** |
| 1 | Mean thickness of caudalanteriorcingulate left hemisphere | 0.01±0.008 | 0.016±0.002 |
| 2 | Mean thickness of pericalcarine right hemisphere | 0.01±0.008 | 0.016±0.002 |
| 3 | Mean thickness of superiorfrontal left hemisphere | 0.01±0.008 | 0.016±0.002 |
| 4 | Mean thickness of lateraloccipital left hemisphere | 0.01±0.008 | 0.016±0.002 |
| 5 | Mean thickness of parahippocampal right hemisphere | 0.009±0.008 | 0.016±0.002 |
| 6 | Mean thickness of medialorbitofrontal right hemisphere | 0.009±0.008 | 0.016±0.002 |
| 7 | Mean thickness of lateraloccipital right hemisphere | 0.009±0.008 | 0.015±0.002 |
| 8 | Mean thickness of rostralmiddlefrontal right hemisphere | 0.009±0.008 | 0.015±0.002 |
| 9 | Mean thickness of isthmuscingulate right hemisphere | 0.009±0.008 | 0.015±0.002 |
| 10 | Mean thickness of paracentral right hemisphere | 0.009±0.008 | 0.015±0.002 |
| 11 | Mean thickness of medialorbitofrontal left hemisphere | 0.009±0.008 | 0.015±0.002 |
| 12 | Mean thickness of superiorfrontal right hemisphere | 0.009±0.008 | 0.015±0.002 |
| 13 | Mean thickness of lateralorbitofrontal right hemisphere | 0.008±0.007 | 0.013±0.002 |
| 14 | Mean thickness of pericalcarine left hemisphere | 0.007±0.007 | 0.013±0.002 |
| 15 | Mean thickness of rostralanteriorcingulate left hemisphere | 0.007±0.007 | 0.012±0.002 |
| 16 | Mean thickness of isthmuscingulate left hemisphere | 0.007±0.007 | 0.012±0.002 |
| 17 | Mean thickness of inferiortemporal right hemisphere | 0.006±0.007 | 0.012±0.002 |
| 18 | Mean thickness of posteriorcingulate left hemisphere | 0.006±0.007 | 0.012±0.002 |
| 19 | Mean thickness of lateralorbitofrontal left hemisphere | 0.006±0.007 | 0.011±0.002 |
| 20 | Mean thickness of fusiform right hemisphere | 0.006±0.007 | 0.011±0.002 |
| 21 | Mean thickness of inferiortemporal left hemisphere | 0.006±0.007 | 0.011±0.002 |
| 22 | Mean thickness of inferiorparietal left hemisphere | 0.006±0.007 | 0.011±0.002 |
| 23 | Mean thickness of superiorparietal right hemisphere | 0.006±0.007 | 0.01±0.002 |
| 24 | Mean thickness of entorhinal right hemisphere | 0.006±0.006 | 0.01±0.001 |
| 25 | Mean thickness of lingual right hemisphere | 0.006±0.007 | 0.01±0.002 |
| 26 | Mean thickness of posteriorcingulate right hemisphere | 0.006±0.006 | 0.01±0.001 |
| 27 | Mean thickness of fusiform left hemisphere | 0.006±0.005 | 0.01±0.001 |
| 28 | Mean thickness of parstriangularis left hemisphere | 0.006±0.005 | 0.009±0.001 |
| 29 | Mean thickness of inferiorparietal right hemisphere | 0.006±0.005 | 0.009±0.001 |
| 30 | Mean thickness of caudalanteriorcingulate right hemisphere | 0.006±0.005 | 0.009±0.001 |
| 31 | Mean thickness of paracentral left hemisphere | 0.005±0.005 | 0.009±0.001 |
| 32 | Mean thickness of precuneus right hemisphere | 0.005±0.005 | 0.008±0.001 |
| 33 | Mean thickness of cuneus right hemisphere | 0.005±0.005 | 0.008±0.001 |
| 34 | Mean thickness of parahippocampal left hemisphere | 0.005±0.005 | 0.008±0.001 |
| 35 | Mean thickness of superiorparietal left hemisphere | 0.005±0.005 | 0.008±0.001 |
| 36 | Mean thickness of cuneus left hemisphere | 0.005±0.005 | 0.008±0.001 |
| 37 | Mean thickness of rostralmiddlefrontal left hemisphere | 0.005±0.005 | 0.008±0.001 |
| 38 | Mean thickness of rostralanteriorcingulate right hemisphere | 0.005±0.005 | 0.008±0.001 |
| 39 | Mean thickness of parsorbitalis left hemisphere | 0.004±0.004 | 0.006±0.001 |
| 40 | Mean thickness of postcentral right hemisphere | 0.004±0.004 | 0.006±0.001 |
| 41 | Mean thickness of parsorbitalis right hemisphere | 0.004±0.004 | 0.006±0.001 |
| 42 | Mean thickness of middletemporal left hemisphere | 0.004±0.004 | 0.006±0.001 |
| 43 | Mean thickness of middletemporal right hemisphere | 0.004±0.003 | 0.006±0.001 |
| 44 | Mean thickness of supramarginal left hemisphere | 0.004±0.003 | 0.006±0.001 |
| 45 | Mean thickness of precuneus left hemisphere | 0.004±0.003 | 0.005±0.001 |
| 46 | Mean thickness of caudalmiddlefrontal right hemisphere | 0.003±0.003 | 0.005±0.001 |
| 47 | Mean thickness of supramarginal right hemisphere | 0.004±0.003 | 0.005±0.001 |
| 48 | Mean thickness of lingual left hemisphere | 0.003±0.003 | 0.005±0.001 |
| 49 | Mean thickness of postcentral left hemisphere | 0.004±0.003 | 0.005±0.001 |
| 50 | Mean thickness of precentral right hemisphere | 0.003±0.004 | 0.005±0.001 |
| 51 | Mean thickness of entorhinal left hemisphere | 0.003±0.004 | 0.005±0.001 |
| 52 | Mean thickness of transversetemporal right hemisphere | 0.003±0.003 | 0.005±0.001 |
| 53 | Mean thickness of parsopercularis right hemisphere | 0.003±0.003 | 0.005±0.001 |
| 54 | Mean thickness of caudalmiddlefrontal left hemisphere | 0.003±0.003 | 0.005±0.001 |
| 55 | Mean thickness of precentral left hemisphere | 0.003±0.003 | 0.005±0.001 |
| 56 | Mean thickness of parsopercularis left hemisphere | 0.004±0.003 | 0.004±0.001 |
| 57 | Mean thickness of transversetemporal left hemisphere | 0.004±0.003 | 0.004±0.001 |
| 58 | Mean thickness of superiortemporal right hemisphere | 0.003±0.003 | 0.004±0.001 |

**Table A6 Feature selection of intensity dataset when predicting AS**. (P-2) linear regression models were trained (with P equal to the number of features in the dataset) using 5-fold cross validation and at each iteration the feature with the lowest correlation with AS was removed from the regressors. The mean $R^2$ in the train and test set of each model across the folds, as well as the name of the feature dropped at each iteration, is reported in the Table

| **Model Number** | **Features dropped** | **AS** | |
|---|---|---|---|
| | | **Test R2 (mean ± std)** | **Train R2 (mean ± std)** |
| **0** | Mean intensity of CSF whole brain | 0.008±0.007 | 0.014±0.002 |
| 1 | Mean intensity of non WM hypointensities whole brain | 0.008±0.007 | 0.014±0.002 |
| 2 | Mean intensity of CC Mid Posterior whole brain | 0.008±0.007 | 0.014±0.002 |
| 3 | Mean intensity of 5th Ventricle whole brain | 0.008±0.006 | 0.014±0.002 |
| 4 | Mean intensity of Cerebellum White Matter left hemisphere | 0.008±0.006 | 0.014±0.002 |
| 5 | Mean intensity of CC Posterior whole brain | 0.008±0.006 | 0.014±0.002 |
| 6 | Volume of WM hypointensities whole brain | 0.008±0.006 | 0.014±0.002 |
| 7 | Mean intensity of Cerebellum Cortex right hemisphere | 0.009±0.006 | 0.014±0.002 |
| 8 | Mean intensity of Amygdala left hemisphere | 0.009±0.006 | 0.014±0.002 |
| 9 | Mean intensity of WM hypointensities whole brain | 0.009±0.006 | 0.014±0.002 |
| 10 | Mean intensity of Cerebellum Cortex left hemisphere | 0.009±0.006 | 0.014±0.002 |
| **11** | **Mean intensity of Cerebellum White Matter right hemisphere** | **0.009±0.006** | **0.014±0.002** |
| 12 | Volume of non WM hypointensities whole brain | 0.009±0.006 | 0.013±0.002 |
| 13 | Mean intensity of Accumbens area right hemisphere | 0.008±0.006 | 0.013±0.001 |
| 14 | Mean intensity of vessel left hemisphere | 0.008±0.006 | 0.013±0.001 |
| 15 | Mean intensity of Amygdala right hemisphere | 0.008±0.006 | 0.013±0.001 |
| 16 | Mean intensity of vessel right hemisphere | 0.007±0.006 | 0.011±0.001 |
| 17 | Mean intensity of Inf Lat Vent right hemisphere | 0.007±0.006 | 0.011±0.002 |
| 18 | Mean intensity of 4th Ventricle whole brain | 0.007±0.006 | 0.011±0.001 |
| 19 | Mean intensity of Inf Lat Vent left hemisphere | 0.006±0.006 | 0.01±0.001 |
| 20 | Mean intensity of CC Central whole brain | 0.006±0.005 | 0.01±0.001 |
| 21 | Mean intensity of Caudate right hemisphere | 0.006±0.006 | 0.01±0.001 |
| 22 | Mean intensity of Optic Chiasm whole brain | 0.006±0.006 | 0.01±0.001 |
| 23 | Mean intensity of Hippocampus left hemisphere | 0.006±0.006 | 0.01±0.001 |
| 24 | Mean intensity of Lateral Ventricle left hemisphere | 0.006±0.006 | 0.01±0.001 |
| 25 | Mean intensity of 3rd Ventricle whole brain | 0.006±0.006 | 0.01±0.001 |
| 26 | Mean intensity of Hippocampus right hemisphere | 0.006±0.005 | 0.009±0.001 |
| 27 | Mean intensity of Lateral Ventricle right hemisphere | 0.006±0.005 | 0.009±0.001 |
| 28 | Mean intensity of VentralDC right hemisphere | 0.007±0.005 | 0.009±0.001 |
| 29 | Mean intensity of VentralDC left hemisphere | 0.007±0.005 | 0.009±0.001 |
| 30 | Mean intensity of Caudate left hemisphere | 0.006±0.005 | 0.008±0.001 |
| 31 | Mean intensity of Accumbens area left hemisphere | 0.006±0.005 | 0.008±0.001 |
| 32 | Mean intensity of choroid plexus left hemisphere | 0.006±0.005 | 0.008±0.001 |
| 33 | Mean intensity of Putamen right hemisphere | 0.006±0.005 | 0.008±0.001 |
| 34 | Mean intensity of Brain Stem whole brain | 0.006±0.005 | 0.008±0.001 |
| 35 | Mean intensity of Pallidum right hemisphere | 0.006±0.005 | 0.008±0.001 |
| 36 | Mean intensity of CC Anterior whole brain | 0.007±0.005 | 0.008±0.001 |
| 37 | Mean intensity of CC Mid Anterior whole brain | 0.006±0.004 | 0.007±0.001 |
| 38 | Mean intensity of Pallidum left hemisphere | 0.005±0.004 | 0.006±0.001 |
| 39 | Mean intensity of choroid plexus right hemisphere | 0.005±0.004 | 0.006±0.001 |
| 40 | Mean intensity of Putamen left hemisphere | 0.002±0.003 | 0.003±0.001 |
| 41 | Mean intensity of Thalamus Proper left hemisphere | 0.002±0.003 | 0.003±0.001 |

**Table A7 Feature selection of volume dataset when predicting AS**. (P-2) linear regression models were trained (with P equal to the number of features in the dataset) using 5-fold cross validation and at each iteration the feature with the lowest correlation with AS was removed from the regressors. The mean $R^2$ in the train and test set of each model across the folds, as well as the name of the feature dropped at each iteration, is reported in the Table.

| Model Number | Features dropped | AS | |
|---|---|---|---|
| | | Test R2 (mean ± std) | Train R2 (mean ± std) |
| 0 | Volume of grey matter in Crus I Cerebellum vermis | 0.025±0.009 | 0.041±0.002 |
| **1** | **Volume of grey matter in Brain Stem** | **0.025±0.008** | **0.041±0.002** |
| 2 | Volume of grey matter in X Cerebellum vermis | 0.023±0.008 | 0.039±0.002 |
| 3 | Volume of grey matter in Pallidum left | 0.023±0.008 | 0.039±0.002 |
| 4 | Volume of grey matter in Pallidum right | 0.023±0.008 | 0.039±0.002 |
| 5 | Volume of grey matter in Caudate left | 0.023±0.008 | 0.039±0.002 |
| 6 | Volume of grey matter in Inferior Temporal Gyrus posterior division left | 0.023±0.008 | 0.039±0.002 |
| 7 | Volume of grey matter in Inferior Frontal Gyrus pars triangularis left | 0.023±0.008 | 0.038±0.002 |
| 8 | Volume of grey matter in Occipital Pole right | 0.023±0.008 | 0.038±0.002 |
| 9 | Volume of grey matter in Juxtapositional Lobule Cortex right | 0.022±0.008 | 0.038±0.002 |
| 10 | Volume of grey matter in Cingulate Gyrus anterior division left | 0.022±0.007 | 0.038±0.002 |
| 11 | Volume of grey matter in I IV Cerebellum right | 0.022±0.008 | 0.037±0.002 |
| 12 | Volume of grey matter in Putamen left | 0.022±0.008 | 0.037±0.002 |
| 13 | Volume of grey matter in Cuneal Cortex left | 0.022±0.008 | 0.037±0.002 |
| 14 | Volume of grey matter in Caudate right | 0.022±0.008 | 0.037±0.002 |
| 15 | Volume of grey matter in Occipital Pole left | 0.022±0.008 | 0.037±0.002 |
| 16 | Volume of grey matter in Inferior Temporal Gyrus anterior division left | 0.021±0.007 | 0.036±0.002 |
| 17 | Volume of grey matter in Inferior Temporal Gyrus anterior division right | 0.021±0.007 | 0.036±0.002 |
| 18 | Volume of grey matter in Inferior Temporal Gyrus temporooccipital part left | 0.021±0.007 | 0.036±0.002 |
| 19 | Volume of grey matter in Cingulate Gyrus anterior division right | 0.021±0.007 | 0.036±0.002 |
| 20 | Volume of grey matter in I IV Cerebellum left | 0.021±0.007 | 0.036±0.002 |
| 21 | Volume of grey matter in Supracalcarine Cortex left | 0.021±0.007 | 0.035±0.002 |
| 22 | Volume of grey matter in Intracalcarine Cortex left | 0.021±0.007 | 0.035±0.002 |
| 23 | Volume of grey matter in Putamen right | 0.021±0.007 | 0.035±0.002 |
| 24 | Volume of grey matter in Inferior Frontal Gyrus pars opercularis right | 0.021±0.007 | 0.035±0.002 |
| 25 | Volume of grey matter in Inferior Frontal Gyrus pars triangularis right | 0.021±0.007 | 0.035±0.002 |
| 26 | Volume of grey matter in Superior Parietal Lobule left | 0.021±0.007 | 0.035±0.002 |
| 27 | Volume of grey matter in Juxtapositional Lobule Cortex left | 0.021±0.007 | 0.035±0.002 |
| 28 | Volume of grey matter in Heschl s Gyrus includes H1 and H2 left | 0.021±0.007 | 0.034±0.002 |
| 29 | Volume of grey matter in Supramarginal Gyrus anterior division right | 0.021±0.007 | 0.034±0.002 |
| 30 | Volume of grey matter in X Cerebellum right | 0.021±0.007 | 0.034±0.002 |
| 31 | Volume of grey matter in X Cerebellum left | 0.021±0.007 | 0.034±0.002 |
| 32 | Volume of grey matter in Inferior Temporal Gyrus temporooccipital part right | 0.021±0.007 | 0.034±0.002 |
| 33 | Volume of grey matter in Superior Parietal Lobule right | 0.021±0.007 | 0.034±0.002 |
| 34 | Volume of grey matter in Supramarginal Gyrus anterior division left | 0.021±0.007 | 0.034±0.002 |
| 35 | Volume of grey matter in VI Cerebellum vermis | 0.021±0.007 | 0.034±0.002 |
| 36 | Volume of grey matter in Middle Temporal Gyrus temporooccipital part left | 0.021±0.007 | 0.034±0.002 |
| 37 | Volume of grey matter in VIIb Cerebellum vermis | 0.021±0.007 | 0.034±0.002 |
| 38 | Volume of grey matter in Supramarginal Gyrus posterior division left | 0.021±0.006 | 0.034±0.002 |
| 39 | Volume of grey matter in Inferior Temporal Gyrus posterior division right | 0.022±0.006 | 0.034±0.002 |
| 40 | Volume of grey matter in Occipital Fusiform Gyrus left | 0.021±0.007 | 0.033±0.002 |
| 41 | Volume of grey matter in Frontal Operculum Cortex left | 0.021±0.007 | 0.033±0.002 |
| 42 | Volume of grey matter in Middle Temporal Gyrus temporooccipital part right | 0.021±0.006 | 0.033±0.002 |
| 43 | Volume of grey matter in Crus II Cerebellum vermis | 0.021±0.006 | 0.033±0.002 |
| 44 | Volume of grey matter in Inferior Frontal Gyrus pars opercularis left | 0.021±0.006 | 0.033±0.002 |
| 45 | Volume of grey matter in Intracalcarine Cortex right | 0.021±0.006 | 0.033±0.002 |
| 46 | Volume of grey matter in Paracingulate Gyrus right | 0.021±0.006 | 0.033±0.002 |
| 47 | Volume of grey matter in Angular Gyrus right | 0.021±0.006 | 0.032±0.001 |
| 48 | Volume of grey matter in Parietal Operculum Cortex left | 0.021±0.006 | 0.032±0.001 |
| 49 | Volume of grey matter in Thalamus left | 0.021±0.006 | 0.032±0.002 |
| 50 | Volume of grey matter in IX Cerebellum right | 0.021±0.006 | 0.032±0.002 |
| 51 | Volume of grey matter in Cuneal Cortex right | 0.021±0.006 | 0.032±0.002 |
| 52 | Volume of grey matter in Planum Temporale left | 0.021±0.006 | 0.032±0.002 |
| 53 | Volume of grey matter in Parietal Operculum Cortex right | 0.021±0.007 | 0.032±0.002 |
| 54 | Volume of grey matter in Parahippocampal Gyrus posterior division left | 0.021±0.007 | 0.031±0.002 |
| 55 | Volume of grey matter in VIIIb Cerebellum vermis | 0.021±0.007 | 0.031±0.002 |
| 56 | Volume of grey matter in Temporal Occipital Fusiform Cortex left | 0.021±0.006 | 0.031±0.002 |
| 57 | Volume of grey matter in Thalamus right | 0.021±0.006 | 0.031±0.002 |
| 58 | Volume of grey matter in Supramarginal Gyrus posterior division right | 0.021±0.006 | 0.031±0.002 |
| 59 | Volume of grey matter in Angular Gyrus left | 0.021±0.006 | 0.031±0.001 |
| 60 | Volume of grey matter in Frontal Medial Cortex left | 0.021±0.006 | 0.03±0.001 |
| 61 | Volume of grey matter in IX Cerebellum left | 0.02±0.006 | 0.03±0.001 |
| 62 | Volume of grey matter in VIIIb Cerebellum right | 0.02±0.006 | 0.03±0.001 |
| 63 | Volume of grey matter in Frontal Operculum Cortex right | 0.02±0.005 | 0.03±0.001 |
| 64 | Volume of grey matter in Ventral Striatum left | 0.02±0.005 | 0.03±0.001 |
| 65 | Volume of grey matter in Middle Temporal Gyrus anterior division right | 0.02±0.005 | 0.029±0.001 |
| 66 | Volume of grey matter in Parahippocampal Gyrus posterior division right | 0.02±0.005 | 0.029±0.001 |
| 67 | Volume of grey matter in Middle Frontal Gyrus right | 0.02±0.005 | 0.029±0.001 |
| 68 | Volume of grey matter in Supracalcarine Cortex right | 0.02±0.005 | 0.029±0.001 |
| 69 | Volume of grey matter in Crus I Cerebellum right | 0.02±0.005 | 0.029±0.001 |
| 70 | Volume of grey matter in Superior Frontal Gyrus left | 0.02±0.005 | 0.029±0.001 |
| 71 | Volume of grey matter in Paracingulate Gyrus left | 0.02±0.005 | 0.029±0.001 |
| 72 | Volume of grey matter in Crus I Cerebellum left | 0.02±0.005 | 0.029±0.001 |
| 73 | Volume of grey matter in Middle Frontal Gyrus left | 0.02±0.006 | 0.028±0.001 |
| 74 | Volume of grey matter in Occipital Fusiform Gyrus right | 0.02±0.006 | 0.028±0.001 |
| 75 | Volume of grey matter in IX Cerebellum vermis | 0.02±0.006 | 0.028±0.001 |
| 76 | Volume of grey matter in Lateral Occipital Cortex  inferior division left | 0.02±0.006 | 0.028±0.001 |
| 77 | Volume of grey matter in Middle Temporal Gyrus  posterior division right | 0.02±0.006 | 0.028±0.001 |
| 78 | Volume of grey matter in Superior Frontal Gyrus right | 0.02±0.006 | 0.028±0.001 |
| 79 | Volume of grey matter in Temporal Occipital Fusiform Cortex right | 0.021±0.005 | 0.028±0.001 |
| 80 | Volume of grey matter in Precuneous Cortex right | 0.021±0.005 | 0.028±0.001 |
| 81 | Volume of grey matter in Central Opercular Cortex right | 0.021±0.006 | 0.028±0.001 |
| 82 | Volume of grey matter in Lateral Occipital Cortex  superior division right | 0.02±0.005 | 0.027±0.001 |
| 83 | Volume of grey matter in Precuneous Cortex left | 0.02±0.005 | 0.027±0.001 |
| 84 | Volume of grey matter in VIIIa Cerebellum vermis | 0.02±0.006 | 0.027±0.001 |
| 85 | Volume of grey matter in Lateral Occipital Cortex  inferior division right | 0.02±0.006 | 0.027±0.001 |
| 86 | Volume of grey matter in Middle Temporal Gyrus  anterior division left | 0.02±0.005 | 0.027±0.001 |
| 87 | Volume of grey matter in Heschl s Gyrus includes H1 and H2 right | 0.02±0.006 | 0.027±0.001 |
| 88 | Volume of grey matter in Frontal Medial Cortex right | 0.02±0.006 | 0.027±0.001 |
| 89 | Volume of grey matter in V Cerebellum right | 0.02±0.006 | 0.026±0.001 |
| 90 | Volume of grey matter in Frontal Orbital Cortex right | 0.02±0.006 | 0.026±0.001 |
| 91 | Volume of grey matter in Temporal Fusiform Cortex  posterior division left | 0.02±0.006 | 0.026±0.002 |
| 92 | Volume of grey matter in Middle Temporal Gyrus  posterior division left | 0.02±0.006 | 0.026±0.001 |
| 93 | Volume of grey matter in Postcentral Gyrus right | 0.02±0.006 | 0.026±0.001 |
| 94 | Volume of grey matter in VIIIb Cerebellum left | 0.02±0.006 | 0.026±0.001 |
| 95 | Volume of grey matter in V Cerebellum left | 0.02±0.006 | 0.026±0.001 |
| 96 | Volume of grey matter in Postcentral Gyrus left | 0.02±0.006 | 0.026±0.001 |
| 97 | Volume of grey matter in Temporal Fusiform Cortex  anterior division right | 0.02±0.006 | 0.025±0.001 |
| 98 | Volume of grey matter in Superior Temporal Gyrus  anterior division right | 0.02±0.006 | 0.025±0.001 |
| 99 | Volume of grey matter in Ventral Striatum right | 0.02±0.006 | 0.025±0.001 |
| 100 | Volume of grey matter in Precentral Gyrus right | 0.02±0.006 | 0.025±0.001 |
| 101 | Volume of grey matter in Lingual Gyrus left | 0.02±0.006 | 0.025±0.001 |
| 102 | Volume of grey matter in Subcallosal Cortex right | 0.02±0.006 | 0.025±0.002 |
| 103 | Volume of grey matter in Planum Polare right | 0.02±0.006 | 0.025±0.002 |
| 104 | Volume of grey matter in Cingulate Gyrus  posterior division left | 0.02±0.006 | 0.025±0.002 |
| 105 | Volume of grey matter in Cingulate Gyrus  posterior division right | 0.02±0.006 | 0.025±0.002 |
| 106 | Volume of grey matter in Superior Temporal Gyrus  posterior division left | 0.02±0.006 | 0.025±0.002 |
| 107 | Volume of grey matter in Subcallosal Cortex left | 0.02±0.006 | 0.025±0.002 |
| 108 | Volume of grey matter in Parahippocampal Gyrus  anterior division right | 0.02±0.006 | 0.024±0.002 |
| 109 | Volume of grey matter in Temporal Fusiform Cortex  posterior division right | 0.02±0.006 | 0.024±0.002 |
| 110 | Volume of grey matter in Frontal Orbital Cortex left | 0.02±0.006 | 0.024±0.002 |
| 111 | Volume of grey matter in Lateral Occipital Cortex  superior division left | 0.02±0.006 | 0.024±0.002 |
| 112 | Volume of grey matter in Lingual Gyrus right | 0.02±0.006 | 0.024±0.002 |
| 113 | Volume of grey matter in Precentral Gyrus left | 0.02±0.006 | 0.024±0.002 |
| 114 | Volume of grey matter in VIIIa Cerebellum left | 0.02±0.007 | 0.024±0.002 |
| 115 | Volume of grey matter in Superior Temporal Gyrus  posterior division right | 0.021±0.007 | 0.024±0.002 |
| 116 | Volume of grey matter in Parahippocampal Gyrus  anterior division left | 0.021±0.007 | 0.024±0.002 |
| 117 | Volume of grey matter in Planum Temporale right | 0.021±0.007 | 0.024±0.002 |
| 118 | Volume of grey matter in VIIIa Cerebellum right | 0.021±0.007 | 0.024±0.002 |
| 119 | Volume of grey matter in Temporal Pole left | 0.021±0.007 | 0.024±0.002 |
| 120 | Volume of grey matter in Temporal Pole right | 0.021±0.007 | 0.024±0.002 |
| 121 | Volume of grey matter in Planum Polare left | 0.021±0.007 | 0.024±0.002 |
| 122 | Volume of grey matter in Central Opercular Cortex left | 0.021±0.007 | 0.024±0.002 |
| 123 | Volume of grey matter in VIIb Cerebellum left | 0.021±0.007 | 0.023±0.002 |
| 124 | Volume of grey matter in Temporal Fusiform Cortex  anterior division left | 0.021±0.007 | 0.023±0.002 |
| 125 | Volume of grey matter in Insular Cortex left | 0.021±0.008 | 0.023±0.002 |
| 126 | Volume of grey matter in Superior Temporal Gyrus  anterior division left | 0.021±0.008 | 0.022±0.002 |
| 127 | Volume of grey matter in VI Cerebellum left | 0.019±0.007 | 0.021±0.002 |
| 128 | Volume of grey matter in Crus II Cerebellum left | 0.019±0.007 | 0.021±0.002 |
| 129 | Volume of grey matter in VI Cerebellum right | 0.02±0.007 | 0.021±0.002 |
| 130 | Volume of grey matter in Crus II Cerebellum right | 0.019±0.007 | 0.021±0.002 |
| 131 | Volume of grey matter in Frontal Pole left | 0.019±0.007 | 0.02±0.002 |
| 132 | Volume of grey matter in VIIb Cerebellum right | 0.019±0.007 | 0.02±0.002 |
| 133 | Volume of grey matter in Insular Cortex right | 0.017±0.005 | 0.018±0.001 |
| 134 | Volume of grey matter in Hippocampus right | 0.017±0.005 | 0.018±0.001 |
| 135 | Volume of grey matter in Frontal Pole right | 0.017±0.005 | 0.018±0.001 |
| 136 | Volume of grey matter in Hippocampus left | 0.016±0.005 | 0.016±0.001 |
| 137 | Volume of grey matter in Amygdala right | 0.025±0.009 | 0.041±0.002 |

**Table A8 Feature selection of FA dataset when predicting DT**. (P-2) linear regression models were trained (with P equal to the number of features in the dataset) using 5-fold cross validation and at each iteration the feature with the lowest correlation with DT was removed from the regressors. The mean $R^2$ in the train and test set of each model across the folds, as well as the name of the feature dropped at each iteration, is reported in the Table.

| Model Number | Features dropped | DT | |
|---|---|---|---|
| | | **Test R2 (mean ± std)** | **Train R2 (mean ± std)** |
| 0 | FA in tract superior thalamic radiation left | 0.0±0.002 | 0.004±0.0 |
| 1 | FA in tract superior thalamic radiation right | 0.0±0.002 | 0.004±0.0 |
| 2 | FA in tract corticospinal tract left | 0.001±0.002 | 0.004±0.001 |
| 3 | FA in tract posterior thalamic radiation right | 0.001±0.002 | 0.004±0.0 |
| 4 | FA in tract corticospinal tract right | 0.001±0.002 | 0.004±0.0 |
| 5 | FA in tract inferior longitudinal fasciculus left | 0.001±0.002 | 0.004±0.0 |
| 6 | FA in tract superior longitudinal fasciculus left | 0.001±0.002 | 0.004±0.0 |
| 7 | FA in tract parahippocampal part of cingulum left | 0.001±0.002 | 0.004±0.0 |
| 8 | FA in tract middle cerebellar peduncle | 0.001±0.002 | 0.003±0.0 |
| 9 | FA in tract inferior fronto occipital fasciculus right | 0.001±0.002 | 0.003±0.0 |
| 10 | FA in tract inferior fronto occipital fasciculus left | 0.001±0.002 | 0.003±0.0 |
| 11 | FA in tract anterior thalamic radiation left | 0.001±0.002 | 0.003±0.0 |
| **12** | **FA in tract forceps minor** | **0.001±0.002** | **0.003±0.0** |
| 13 | FA in tract posterior thalamic radiation left | 0.001±0.002 | 0.003±0.0 |
| 14 | FA in tract parahippocampal part of cingulum right | 0.001±0.001 | 0.002±0.0 |
| 15 | FA in tract inferior longitudinal fasciculus right | 0.001±0.001 | 0.002±0.0 |
| 16 | FA in tract medial lemniscus left | 0.001±0.001 | 0.002±0.0 |
| 17 | FA in tract anterior thalamic radiation right | 0.001±0.001 | 0.002±0.0 |
| 18 | FA in tract forceps major | 0.001±0.001 | 0.002±0.0 |
| 19 | FA in tract superior longitudinal fasciculus right | 0.001±0.001 | 0.002±0.0 |
| 20 | FA in tract acoustic radiation left | 0.001±0.001 | 0.002±0.0 |
| 21 | FA in tract medial lemniscus right | 0.001±0.001 | 0.002±0.0 |
| 22 | FA in tract cingulate gyrus part of cingulum left | 0.001±0.0 | 0.001±0.0 |
| 23 | FA in tract cingulate gyrus part of cingulum right | 0.001±0.0 | 0.001±0.0 |
| 24 | FA in tract acoustic radiation right | 0.001±0.0 | 0.001±0.0 |

**Table A9 Feature selection of thickness dataset when predicting DT**. (P-2) linear regression models were trained (with P equal to the number of features in the dataset) using 5-fold cross validation and at each iteration the feature with the lowest correlation with DT was removed from the regressors. The mean $R^2$ in the train and test set of each model across the folds, as well as the name of the feature dropped at each iteration, is reported in the Table.

| Model Number | Features dropped | DT | |
|---|---|---|---|
| | | **Test R2 (mean ± std)** | **Train R2 (mean ± std)** |
| **0** | Mean thickness of postcentral left hemisphere | -0.001±0.002 | 0.007±0.001 |
| 1 | Mean thickness of isthmuscingulate left hemisphere | -0.001±0.003 | 0.007±0.001 |
| 2 | Mean thickness of entorhinal right hemisphere | -0.0±0.003 | 0.007±0.001 |
| 3 | Mean thickness of postcentral right hemisphere | -0.001±0.003 | 0.007±0.001 |
| 4 | Mean thickness of superiorparietal left hemisphere | -0.0±0.003 | 0.007±0.001 |
| 5 | Mean thickness of isthmuscingulate right hemisphere | -0.0±0.003 | 0.007±0.001 |
| 6 | Mean thickness of paracentral right hemisphere | -0.0±0.003 | 0.007±0.001 |
| 7 | Mean thickness of parstriangularis right hemisphere | -0.0±0.003 | 0.007±0.001 |
| 8 | Mean thickness of superiorparietal right hemisphere | -0.0±0.003 | 0.007±0.001 |
| 9 | Mean thickness of cuneus left hemisphere | -0.001±0.003 | 0.006±0.001 |
| 10 | Mean thickness of rostralmiddlefrontal right hemisphere | -0.001±0.003 | 0.006±0.001 |
| 11 | Mean thickness of posteriorcingulate right hemisphere | -0.001±0.003 | 0.006±0.001 |
| 12 | Mean thickness of parsopercularis left hemisphere | -0.001±0.003 | 0.006±0.001 |
| 13 | Mean thickness of parstriangularis left hemisphere | -0.0±0.003 | 0.006±0.001 |
| 14 | Mean thickness of rostralanteriorcingulate right hemisphere | -0.0±0.003 | 0.006±0.001 |
| 15 | Mean thickness of parsopercularis right hemisphere | -0.0±0.003 | 0.006±0.001 |
| 16 | Mean thickness of posteriorcingulate left hemisphere | -0.0±0.003 | 0.006±0.001 |
| 17 | Mean thickness of lateraloccipital right hemisphere | -0.0±0.003 | 0.006±0.001 |
| 18 | Mean thickness of precuneus right hemisphere | -0.0±0.003 | 0.005±0.001 |
| 19 | Mean thickness of lateraloccipital left hemisphere | -0.0±0.003 | 0.005±0.001 |
| 20 | Mean thickness of precentral right hemisphere | -0.001±0.003 | 0.005±0.001 |
| 21 | Mean thickness of precentral left hemisphere | -0.0±0.003 | 0.005±0.001 |
| 22 | Mean thickness of rostralmiddlefrontal left hemisphere | -0.0±0.003 | 0.005±0.001 |
| 23 | Mean thickness of caudalmiddlefrontal right hemisphere | -0.0±0.003 | 0.005±0.001 |
| 24 | Mean thickness of entorhinal left hemisphere | -0.0±0.003 | 0.005±0.001 |
| 25 | Mean thickness of medialorbitofrontal right hemisphere | 0.0±0.002 | 0.005±0.001 |
| 26 | Mean thickness of caudalmiddlefrontal left hemisphere | 0.0±0.003 | 0.005±0.001 |
| 27 | Mean thickness of pericalcarine right hemisphere | 0.0±0.003 | 0.005±0.001 |
| 28 | Mean thickness of transversetemporal left hemisphere | 0.001±0.003 | 0.005±0.001 |
| 29 | Mean thickness of parsorbitalis right hemisphere | 0.001±0.003 | 0.005±0.001 |
| 30 | Mean thickness of supramarginal right hemisphere | 0.001±0.003 | 0.005±0.001 |
| 31 | Mean thickness of lingual right hemisphere | 0.001±0.003 | 0.005±0.001 |
| 32 | Mean thickness of fusiform left hemisphere | 0.001±0.003 | 0.005±0.001 |
| 33 | Mean thickness of inferiorparietal right hemisphere | 0.001±0.003 | 0.005±0.001 |
| 34 | Mean thickness of paracentral left hemisphere | 0.001±0.003 | 0.004±0.001 |
| 35 | Mean thickness of precuneus left hemisphere | 0.001±0.003 | 0.004±0.001 |
| 36 | Mean thickness of transversetemporal right hemisphere | 0.001±0.003 | 0.004±0.001 |
| 37 | Mean thickness of medialorbitofrontal left hemisphere | 0.001±0.003 | 0.004±0.001 |
| 38 | Mean thickness of lingual left hemisphere | 0.001±0.003 | 0.004±0.001 |
| 39 | Mean thickness of inferiorparietal left hemisphere | 0.001±0.002 | 0.004±0.001 |
| 40 | Mean thickness of parahippocampal left hemisphere | 0.001±0.002 | 0.004±0.001 |
| 41 | Mean thickness of pericalcarine left hemisphere | 0.001±0.002 | 0.004±0.001 |
| 42 | Mean thickness of superiorfrontal right hemisphere | 0.001±0.002 | 0.004±0.001 |
| 43 | Mean thickness of supramarginal left hemisphere | 0.001±0.002 | 0.004±0.001 |
| 44 | Mean thickness of lateralorbitofrontal left hemisphere | 0.001±0.002 | 0.004±0.001 |
| **45** | **Mean thickness of cuneus right hemisphere** | **0.001±0.003** | **0.004±0.001** |
| 46 | Mean thickness of superiorfrontal left hemisphere | 0.0±0.003 | 0.002±0.001 |
| 47 | Mean thickness of parahippocampal right hemisphere | 0.0±0.002 | 0.002±0.001 |
| 48 | Mean thickness of parsorbitalis left hemisphere | 0.0±0.002 | 0.002±0.001 |
| 49 | Mean thickness of caudalanteriorcingulate left hemisphere | 0.0±0.002 | 0.002±0.001 |
| 50 | Mean thickness of fusiform right hemisphere | 0.001±0.002 | 0.002±0.001 |
| 51 | Mean thickness of rostralanteriorcingulate left hemisphere | 0.001±0.002 | 0.002±0.001 |
| 52 | Mean thickness of superiortemporal right hemisphere | 0.001±0.002 | 0.002±0.0 |
| 53 | Mean thickness of caudalanteriorcingulate right hemisphere | 0.001±0.002 | 0.002±0.0 |
| 54 | Mean thickness of superiortemporal left hemisphere | 0.001±0.002 | 0.002±0.0 |
| 55 | Mean thickness of middletemporal right hemisphere | 0.001±0.002 | 0.002±0.001 |
| 56 | Mean thickness of lateralorbitofrontal right hemisphere | 0.001±0.002 | 0.002±0.001 |
| 57 | Mean thickness of inferiortemporal right hemisphere | 0.001±0.002 | 0.002±0.0 |
| 58 | Mean thickness of inferiortemporal left hemisphere | 0.001±0.002 | 0.001±0.0 |

**Table A10 Feature selection of intensity dataset when predicting DT**. (P-2) linear regression models were trained (with P equal to the number of features in the dataset) using 5-fold cross validation and at each iteration the feature with the lowest correlation with DT was removed from the regressors. The mean $R^2$ in the train and test set of each model across the folds, as well as the name of the feature dropped at each iteration, is reported in the Table.

| Model Number | Features dropped | DT | |
|---|---|---|---|
| | | **Test R2 (mean ± std)** | **Train R2 (mean ± std)** |
| **0** | Mean intensity of Accumbens area right hemisphere | -0.0±0.002 | 0.005±0.001 |
| 1 | Mean intensity of Cerebellum White Matter right hemisphere | -0.001±0.002 | 0.005±0.001 |
| 2 | Mean intensity of Caudate left hemisphere | -0.001±0.002 | 0.005±0.001 |
| 3 | Mean intensity of Caudate right hemisphere | -0.001±0.002 | 0.005±0.001 |
| 4 | Mean intensity of Inf Lat Vent left hemisphere | -0.001±0.003 | 0.004±0.001 |
| 5 | Mean intensity of Inf Lat Vent right hemisphere | -0.001±0.002 | 0.004±0.001 |
| 6 | Mean intensity of WM hypointensities whole brain | -0.001±0.002 | 0.004±0.0 |
| 7 | Mean intensity of Putamen left hemisphere | -0.001±0.001 | 0.004±0.0 |
| 8 | Mean intensity of 5th Ventricle whole brain | -0.001±0.001 | 0.004±0.0 |
| 9 | Mean intensity of vessel right hemisphere | -0.001±0.001 | 0.004±0.0 |
| 10 | Mean intensity of choroid plexus right hemisphere | -0.0±0.001 | 0.004±0.0 |
| **11** | Mean intensity of CC Mid Anterior whole brain | -0.0±0.001 | 0.004±0.0 |
| 12 | Mean intensity of Pallidum right hemisphere | -0.0±0.001 | 0.004±0.0 |
| 13 | Mean intensity of non WM hypointensities whole brain | -0.0±0.001 | 0.004±0.0 |
| 14 | Mean intensity of choroid plexus left hemisphere | -0.0±0.001 | 0.004±0.0 |
| 15 | Volume of non WM hypointensities whole brain | -0.0±0.001 | 0.004±0.0 |
| 16 | Mean intensity of Putamen right hemisphere | -0.0±0.001 | 0.003±0.0 |
| 17 | Mean intensity of Cerebellum White Matter left hemisphere | -0.001±0.001 | 0.003±0.0 |
| 18 | Mean intensity of Pallidum left hemisphere | -0.001±0.001 | 0.003±0.0 |
| 19 | Mean intensity of CC Posterior whole brain | -0.001±0.001 | 0.003±0.0 |
| 20 | Mean intensity of Optic Chiasm whole brain | -0.0±0.001 | 0.003±0.0 |
| 21 | Mean intensity of CC Anterior whole brain | -0.0±0.001 | 0.003±0.0 |
| 22 | Mean intensity of Thalamus Proper left hemisphere | -0.0±0.001 | 0.003±0.0 |
| 23 | Mean intensity of vessel left hemisphere | -0.001±0.001 | 0.003±0.0 |
| 24 | Volume of WM hypointensities whole brain | -0.0±0.001 | 0.002±0.0 |
| 25 | Mean intensity of VentralDC left hemisphere | -0.0±0.001 | 0.002±0.0 |
| 26 | Mean intensity of Amygdala right hemisphere | 0.0±0.001 | 0.002±0.0 |
| 27 | Mean intensity of Brain Stem whole brain | 0.0±0.001 | 0.002±0.0 |
| 28 | Mean intensity of VentralDC right hemisphere | 0.0±0.001 | 0.002±0.0 |
| 29 | Mean intensity of Thalamus Proper right hemisphere | 0.0±0.001 | 0.002±0.0 |
| 30 | Mean intensity of CC Central whole brain | 0.0±0.001 | 0.002±0.0 |
| 31 | Mean intensity of CC Mid Posterior whole brain | 0.0±0.001 | 0.002±0.0 |
| 32 | Mean intensity of Cerebellum Cortex right hemisphere | 0.0±0.001 | 0.002±0.0 |
| 33 | Mean intensity of Accumbens area left hemisphere | -0.0±0.001 | 0.001±0.0 |
| 34 | Mean intensity of Lateral Ventricle left hemisphere | 0.0±0.001 | 0.001±0.0 |
| 35 | Mean intensity of Lateral Ventricle right hemisphere | 0.0±0.001 | 0.001±0.0 |
| 36 | Mean intensity of Amygdala left hemisphere | 0.0±0.001 | 0.001±0.0 |
| 37 | Mean intensity of CSF whole brain | 0.001±0.001 | 0.001±0.0 |
| 38 | Mean intensity of Cerebellum Cortex left hemisphere | 0.001±0.001 | 0.001±0.0 |
| 39 | Mean intensity of 3rd Ventricle whole brain | 0.001±0.001 | 0.001±0.0 |
| 40 | Mean intensity of Hippocampus left hemisphere | 0.001±0.001 | 0.001±0.0 |
| **41** | **Mean intensity of 4th Ventricle whole brain** | **0.001±0.001** | **0.001±0.0** |

**Table A11 Feature selection of volume dataset when predicting AS**. (P-2) linear regression models were trained (with P equal to the number of features in the dataset) using 5-fold cross validation and at each iteration the feature with the lowest correlation with DT was removed from the regressors. The mean $R^2$ in the train and test set of each model across the folds, as well as the name of the feature dropped at each iteration, is reported in the Table.

| **Model Number** | **Features dropped** | **DT** | |
|---|---|---|---|
| | | **Test R2 (mean ± std)** | **Train R2 (mean ± std)** |
| 0 | Volume of grey matter in Hippocampus left | -0.003±0.004 | 0.015±0.001 |
| 1 | Volume of grey matter in Planum Polare right | -0.002±0.004 | 0.015±0.001 |
| 2 | Volume of grey matter in Inferior Temporal Gyrus temporooccipital part right | -0.002±0.005 | 0.015±0.001 |
| 3 | Volume of grey matter in Lateral Occipital Cortex inferior division left | -0.002±0.004 | 0.015±0.001 |
| 4 | Volume of grey matter in Inferior Frontal Gyrus pars triangularis right | -0.002±0.004 | 0.015±0.001 |
| 5 | Volume of grey matter in Caudate right | -0.002±0.004 | 0.015±0.001 |
| 6 | Volume of grey matter in Middle Temporal Gyrus anterior division left | -0.002±0.004 | 0.015±0.001 |
| 7 | Volume of grey matter in Temporal Fusiform Cortex posterior division left | -0.002±0.004 | 0.015±0.001 |
| 8 | Volume of grey matter in Planum Temporale left | -0.002±0.004 | 0.014±0.001 |
| 9 | Volume of grey matter in Cingulate Gyrus anterior division left | -0.002±0.004 | 0.014±0.001 |
| 10 | Volume of grey matter in Supramarginal Gyrus anterior division left | -0.002±0.004 | 0.014±0.001 |
| 11 | Volume of grey matter in Caudate left | -0.002±0.004 | 0.014±0.001 |
| 12 | Volume of grey matter in Cingulate Gyrus posterior division left | -0.002±0.004 | 0.014±0.001 |
| 13 | Volume of grey matter in Supramarginal Gyrus posterior division right | -0.002±0.004 | 0.014±0.001 |
| 14 | Volume of grey matter in Supramarginal Gyrus anterior division right | -0.002±0.004 | 0.014±0.001 |
| 15 | Volume of grey matter in Middle Temporal Gyrus anterior division right | -0.002±0.004 | 0.014±0.001 |
| 16 | Volume of grey matter in Subcallosal Cortex left | -0.002±0.004 | 0.014±0.001 |
| 17 | Volume of grey matter in Superior Temporal Gyrus anterior division left | -0.002±0.004 | 0.014±0.001 |
| 18 | Volume of grey matter in Planum Polare left | -0.002±0.004 | 0.014±0.001 |
| 19 | Volume of grey matter in Juxtapositional Lobule Cortex right | -0.002±0.004 | 0.014±0.001 |
| 20 | Volume of grey matter in Lateral Occipital Cortex inferior division right | -0.001±0.004 | 0.014±0.001 |
| 21 | Volume of grey matter in Inferior Frontal Gyrus pars opercularis left | -0.001±0.004 | 0.014±0.001 |
| 22 | Volume of grey matter in I IV Cerebellum left | -0.001±0.004 | 0.014±0.001 |
| 23 | Volume of grey matter in Thalamus left | -0.001±0.004 | 0.014±0.001 |
| 24 | Volume of grey matter in Frontal Pole right | -0.001±0.004 | 0.014±0.001 |
| 25 | Volume of grey matter in Precentral Gyrus left | -0.002±0.004 | 0.013±0.001 |
| 26 | Volume of grey matter in Inferior Frontal Gyrus pars opercularis right | -0.001±0.004 | 0.013±0.001 |
| 27 | Volume of grey matter in Hippocampus right | -0.001±0.004 | 0.013±0.001 |
| 28 | Volume of grey matter in Central Opercular Cortex right | -0.001±0.004 | 0.013±0.001 |
| 29 | Volume of grey matter in Paracingulate Gyrus right | -0.001±0.004 | 0.013±0.001 |
| 30 | Volume of grey matter in Temporal Fusiform Cortex posterior division right | -0.001±0.004 | 0.013±0.001 |
| 31 | Volume of grey matter in Frontal Operculum Cortex left | -0.001±0.004 | 0.013±0.001 |
| 32 | Volume of grey matter in Superior Temporal Gyrus anterior division right | -0.001±0.004 | 0.013±0.001 |
| 33 | Volume of grey matter in Inferior Temporal Gyrus posterior division left | -0.001±0.004 | 0.013±0.001 |
| 34 | Volume of grey matter in Parahippocampal Gyrus posterior division left | -0.001±0.004 | 0.013±0.001 |
| 35 | Volume of grey matter in Heschl s Gyrus includes H1 and H2 left | -0.001±0.004 | 0.013±0.001 |
| 36 | Volume of grey matter in Inferior Temporal Gyrus anterior division left | -0.001±0.004 | 0.013±0.001 |
| 37 | Volume of grey matter in Superior Temporal Gyrus posterior division left | -0.001±0.004 | 0.012±0.001 |
| 38 | Volume of grey matter in Lingual Gyrus left | -0.001±0.004 | 0.012±0.001 |
| 39 | Volume of grey matter in Frontal Medial Cortex left | -0.001±0.004 | 0.012±0.001 |
| 40 | Volume of grey matter in Middle Temporal Gyrus posterior division right | -0.001±0.004 | 0.012±0.001 |
| 41 | Volume of grey matter in Inferior Frontal Gyrus pars triangularis left | -0.001±0.003 | 0.012±0.001 |
| 42 | Volume of grey matter in Angular Gyrus right | -0.001±0.003 | 0.012±0.001 |
| 43 | Volume of grey matter in Temporal Pole left | -0.0±0.003 | 0.012±0.001 |
| 44 | Volume of grey matter in Frontal Pole left | -0.0±0.003 | 0.012±0.001 |
| 45 | Volume of grey matter in Inferior Temporal Gyrus anterior division right | -0.0±0.004 | 0.012±0.001 |
| 46 | Volume of grey matter in Planum Temporale right | 0.0±0.004 | 0.012±0.001 |
| 47 | Volume of grey matter in X Cerebellum vermis | 0.0±0.004 | 0.012±0.001 |
| 48 | Volume of grey matter in Cuneal Cortex left | 0.0±0.004 | 0.012±0.001 |
| 49 | Volume of grey matter in Middle Temporal Gyrus posterior division left | 0.0±0.004 | 0.012±0.001 |
| 50 | Volume of grey matter in I IV Cerebellum right | 0.0±0.004 | 0.012±0.001 |
| 51 | Volume of grey matter in Angular Gyrus left | 0.0±0.004 | 0.011±0.001 |
| 52 | Volume of grey matter in Inferior Temporal Gyrus posterior division right | 0.0±0.004 | 0.011±0.001 |
| 53 | Volume of grey matter in Temporal Occipital Fusiform Cortex right | 0.0±0.003 | 0.011±0.001 |
| 54 | Volume of grey matter in Crus I Cerebellum vermis | 0.0±0.004 | 0.011±0.001 |
| 55 | Volume of grey matter in Cingulate Gyrus posterior division right | 0.0±0.003 | 0.011±0.001 |
| 56 | Volume of grey matter in Juxtapositional Lobule Cortex left | 0.0±0.003 | 0.011±0.001 |
| 57 | Volume of grey matter in Middle Temporal Gyrus temporooccipital part right | 0.0±0.003 | 0.011±0.001 |
| 58 | Volume of grey matter in Subcallosal Cortex right | 0.001±0.003 | 0.011±0.001 |
| 59 | Volume of grey matter in VIIb Cerebellum vermis | 0.001±0.003 | 0.011±0.001 |
| 60 | Volume of grey matter in Precentral Gyrus right | 0.001±0.003 | 0.011±0.001 |
| 61 | Volume of grey matter in Lingual Gyrus right | 0.001±0.003 | 0.011±0.001 |
| 62 | Volume of grey matter in Parahippocampal Gyrus posterior division right | 0.001±0.003 | 0.011±0.001 |
| 63 | Volume of grey matter in Insular Cortex right | 0.001±0.003 | 0.011±0.001 |
| 64 | Volume of grey matter in Frontal Operculum Cortex right | 0.001±0.003 | 0.011±0.001 |
| 65 | Volume of grey matter in Central Opercular Cortex left | 0.001±0.003 | 0.011±0.001 |
| 66 | Volume of grey matter in Superior Frontal Gyrus right | 0.001±0.003 | 0.011±0.001 |
| 67 | Volume of grey matter in Heschl s Gyrus includes H1 and H2 right | 0.001±0.003 | 0.011±0.001 |
| 68 | Volume of grey matter in Thalamus right | 0.001±0.003 | 0.01±0.001 |
| 69 | Volume of grey matter in Superior Parietal Lobule right | 0.001±0.003 | 0.01±0.001 |
| 70 | Volume of grey matter in Cingulate Gyrus  anterior division right | 0.001±0.003 | 0.01±0.001 |
| 71 | Volume of grey matter in VI Cerebellum right | 0.001±0.003 | 0.01±0.001 |
| 72 | Volume of grey matter in Pallidum right | 0.001±0.003 | 0.01±0.001 |
| 73 | Volume of grey matter in Postcentral Gyrus left | 0.001±0.004 | 0.01±0.001 |
| 74 | Volume of grey matter in Insular Cortex left | 0.001±0.004 | 0.01±0.001 |
| 75 | Volume of grey matter in Parietal Operculum Cortex left | 0.001±0.004 | 0.01±0.001 |
| 76 | Volume of grey matter in VI Cerebellum left | 0.001±0.004 | 0.01±0.001 |
| 77 | Volume of grey matter in Pallidum left | 0.001±0.004 | 0.01±0.001 |
| 78 | Volume of grey matter in Supracalcarine Cortex right | 0.001±0.004 | 0.01±0.001 |
| 79 | Volume of grey matter in Occipital Fusiform Gyrus right | 0.001±0.004 | 0.01±0.001 |
| 80 | Volume of grey matter in Superior Parietal Lobule left | 0.002±0.004 | 0.01±0.001 |
| 81 | Volume of grey matter in Temporal Pole right | 0.002±0.004 | 0.01±0.001 |
| 82 | Volume of grey matter in X Cerebellum right | 0.002±0.004 | 0.01±0.001 |
| 83 | Volume of grey matter in Brain Stem | 0.002±0.004 | 0.01±0.001 |
| 84 | Volume of grey matter in Inferior Temporal Gyrus  temporooccipital part left | 0.002±0.004 | 0.009±0.001 |
| 85 | Volume of grey matter in Temporal Occipital Fusiform Cortex left | 0.002±0.004 | 0.009±0.001 |
| 86 | Volume of grey matter in Amygdala right | 0.002±0.004 | 0.009±0.001 |
| 87 | Volume of grey matter in Middle Frontal Gyrus left | 0.002±0.004 | 0.009±0.001 |
| 88 | Volume of grey matter in Postcentral Gyrus right | 0.002±0.004 | 0.009±0.001 |
| 89 | Volume of grey matter in Frontal Orbital Cortex left | 0.002±0.003 | 0.009±0.001 |
| 90 | Volume of grey matter in Amygdala left | 0.002±0.004 | 0.009±0.001 |
| 91 | Volume of grey matter in V Cerebellum left | 0.002±0.004 | 0.009±0.001 |
| 92 | Volume of grey matter in Middle Frontal Gyrus right | 0.002±0.004 | 0.008±0.001 |
| 93 | Volume of grey matter in Superior Frontal Gyrus left | 0.002±0.004 | 0.008±0.001 |
| 94 | Volume of grey matter in Superior Temporal Gyrus  posterior division right | 0.002±0.004 | 0.008±0.001 |
| 95 | Volume of grey matter in Parahippocampal Gyrus  anterior division right | 0.002±0.004 | 0.008±0.001 |
| 96 | Volume of grey matter in Supramarginal Gyrus  posterior division left | 0.002±0.004 | 0.008±0.001 |
| **97** | **Volume of grey matter in Frontal Medial Cortex right** | **0.002±0.004** | **0.008±0.001** |
| 98 | Volume of grey matter in Frontal Orbital Cortex right | 0.002±0.004 | 0.008±0.001 |
| 99 | Volume of grey matter in IX Cerebellum vermis | 0.002±0.004 | 0.007±0.001 |
| 100 | Volume of grey matter in Paracingulate Gyrus left | 0.002±0.004 | 0.007±0.001 |
| 101 | Volume of grey matter in VIIIb Cerebellum vermis | 0.002±0.003 | 0.007±0.001 |
| 102 | Volume of grey matter in V Cerebellum right | 0.002±0.003 | 0.007±0.001 |
| 103 | Volume of grey matter in Lateral Occipital Cortex  superior division right | 0.002±0.003 | 0.007±0.001 |
| 104 | Volume of grey matter in Parahippocampal Gyrus  anterior division left | 0.002±0.003 | 0.007±0.001 |
| 105 | Volume of grey matter in Middle Temporal Gyrus  temporooccipital part left | 0.002±0.003 | 0.007±0.001 |
| 106 | Volume of grey matter in Crus II Cerebellum vermis | 0.002±0.003 | 0.006±0.001 |
| 107 | Volume of grey matter in Supracalcarine Cortex left | 0.002±0.003 | 0.006±0.001 |
| 108 | Volume of grey matter in Temporal Fusiform Cortex  anterior division left | 0.002±0.003 | 0.006±0.001 |
| 109 | Volume of grey matter in Parietal Operculum Cortex right | 0.002±0.003 | 0.006±0.001 |
| 110 | Volume of grey matter in Precuneous Cortex right | 0.002±0.003 | 0.006±0.001 |
| 111 | Volume of grey matter in IX Cerebellum left | 0.002±0.003 | 0.006±0.001 |
| 112 | Volume of grey matter in Temporal Fusiform Cortex  anterior division right | 0.002±0.003 | 0.006±0.001 |
| 113 | Volume of grey matter in X Cerebellum left | 0.002±0.003 | 0.005±0.001 |
| 114 | Volume of grey matter in VI Cerebellum vermis | 0.002±0.003 | 0.005±0.001 |
| 115 | Volume of grey matter in Ventral Striatum left | 0.002±0.003 | 0.005±0.001 |
| 116 | Volume of grey matter in VIIIa Cerebellum vermis | 0.002±0.003 | 0.005±0.001 |
| 117 | Volume of grey matter in Cuneal Cortex right | 0.002±0.003 | 0.005±0.001 |
| 118 | Volume of grey matter in Lateral Occipital Cortex  superior division left | 0.002±0.003 | 0.005±0.001 |
| 119 | Volume of grey matter in VIIIb Cerebellum left | 0.002±0.003 | 0.005±0.001 |
| 120 | Volume of grey matter in Precuneous Cortex left | 0.002±0.003 | 0.005±0.001 |
| 121 | Volume of grey matter in Putamen left | 0.002±0.003 | 0.005±0.001 |
| 122 | Volume of grey matter in Occipital Fusiform Gyrus left | 0.002±0.003 | 0.005±0.001 |
| 123 | Volume of grey matter in Ventral Striatum right | 0.002±0.003 | 0.004±0.001 |
| 124 | Volume of grey matter in IX Cerebellum right | 0.002±0.002 | 0.004±0.001 |
| 125 | Volume of grey matter in Occipital Pole left | 0.002±0.002 | 0.004±0.001 |
| 126 | Volume of grey matter in VIIIb Cerebellum right | 0.002±0.002 | 0.004±0.001 |
| 127 | Volume of grey matter in Intracalcarine Cortex right | 0.002±0.002 | 0.004±0.001 |
| 128 | Volume of grey matter in VIIIa Cerebellum left | 0.002±0.002 | 0.004±0.001 |
| 129 | Volume of grey matter in Putamen right | 0.002±0.002 | 0.004±0.001 |
| 130 | Volume of grey matter in Crus I Cerebellum left | 0.002±0.002 | 0.003±0.0 |
| 131 | Volume of grey matter in Crus I Cerebellum right | 0.002±0.001 | 0.003±0.0 |
| 132 | Volume of grey matter in Occipital Pole right | 0.002±0.001 | 0.003±0.0 |
| 133 | Volume of grey matter in Intracalcarine Cortex left | 0.002±0.001 | 0.003±0.0 |
| 134 | Volume of grey matter in VIIIa Cerebellum right | 0.001±0.001 | 0.002±0.0 |
| 135 | Volume of grey matter in VIIb Cerebellum right | 0.001±0.001 | 0.001±0.0 |
| 136 | Volume of grey matter in VIIb Cerebellum left | 0.001±0.001 | 0.001±0.0 |
| 137 | Volume of grey matter in Crus II Cerebellum left | 0.001±0.001 | 0.001±0.0 |

**Table A12 List of most frequent words in the literature related to the neural correlates of DT extracted following the univariate analysis pipeline.** The frequency of occurrence of each word was calculated across all papers for each individual brain region, and normalized between 0 and 1, with 1 corresponding to the most frequent word. The frequency scores of the 5 words with the highest values were added for all the brain regions related to DT. The neural correlates of DT were extracted following the univariate analysis pipeline.

| Word | Cumulative frequency of occurrence |
|---|---|
| visual | 3.646944 |
| stimulus | 1.530769 |
| lesion | 1.298545 |
| learning | 1.282701 |
| age | 1.174014 |
| schizophrenia | 1.046582 |
| reward | 1.000000 |
| motor | 1.000000 |
| asd | 1.000000 |
| auditory | 1.000000 |
| speech | 0.980220 |
| word | 0.893082 |
| speed | 0.806647 |
| motion | 0.800604 |
| reversal | 0.800448 |
| number | 0.754717 |
| layer | 0.716012 |
| choice | 0.665919 |
| early | 0.557692 |
| field | 0.556604 |
| social | 0.430085 |
| pattern | 0.419872 |
| measure | 0.407051 |

**Table A13 List of most frequent words in the literature related to the neural correlates of DT extracted following the multivariate analysis pipeline.** The frequency of occurrence of each word was calculated across all papers for each individual brain region, and normalized between 0 and 1, with 1 corresponding to the most frequent word. The frequency scores of the 5 words with the highest values were added for all the brain regions related to DT. The neural correlates of DT were extracted following the multivariate analysis pipeline.

| Word | Cumulative frequency of occurrence |
| --- | --- |
| visual | 2.572749 |
| age | 2.000000 |
| motor | 1.800718 |
| lesion | 1.315089 |
| stimulus | 1.273666 |
| object | 1.000000 |
| reward | 1.000000 |
| dyslexia | 0.978723 |
| learning | 0.901345 |
| schizophrenia | 0.872340 |
| fmri | 0.851064 |
| speed | 0.806647 |
| motion | 0.800604 |
| reversal | 0.800448 |
| state | 0.797872 |
| impairment | 0.728905 |
| ad | 0.728905 |
| measure | 0.718133 |
| layer | 0.716012 |
| anisotropy | 0.697761 |
| sensory | 0.679458 |
| injury | 0.679104 |
| brainstem | 0.670429 |
| choice | 0.665919 |
| approach | 0.660448 |
| clinical | 0.604966 |
| stimulation | 0.519187 |
| shape | 0.299484 |
| representation | 0.237522 |

**Table A14 List of most frequent words in the literature related to the neural correlates of AS extracted following the multivariate analysis pipeline.** The frequency of occurrence of each word was calculated across all papers for each individual brain region, and normalized between 0 and 1, with 1 corresponding to the most frequent word. The frequency scores of the 5 words with the highest values were added for all the brain regions related to AS. The neural correlates of AS were extracted following the multivariate analysis pipeline.

| Word | Cumulative frequency of occurrence |
|---|---|
| memory | 3.755957 |
| age | 3.240682 |
| auditory | 2.000000 |
| motor | 1.721180 |
| schizophrenia | 1.408927 |
| asd | 1.000000 |
| reward | 1.000000 |
| error | 1.000000 |
| anxiety | 1.000000 |
| speech | 0.980220 |
| ad | 0.960422 |
| word | 0.893082 |
| behavior | 0.811451 |
| visual | 0.779874 |
| number | 0.754717 |
| dopamine | 0.734452 |
| emotional | 0.704225 |
| emotion | 0.686620 |
| sensory | 0.679458 |
| size | 0.678261 |
| development | 0.653623 |
| surface | 0.638070 |
| splenium | 0.633333 |
| relationship | 0.620053 |
| clinical | 0.604966 |
| hemisphere | 0.604348 |
| lower | 0.603217 |
| shell | 0.589339 |
| conflict | 0.561404 |
| stimulus | 0.530769 |
| stimulation | 0.519187 |
| case | 0.512415 |
| expression | 0.504573 |
| action | 0.501385 |
| microstructural | 0.493404 |
| subject | 0.470330 |
| drug | 0.454097 |
| trial | 0.435826 |
| pattern | 0.429840 |
| synaptic | 0.419207 |
| monitoring | 0.419206 |
| learning | 0.380335 |
| stress | 0.297256 |
| primary | 0.293073 |
| hearing | 0.273535 |
| memory | 3.755957 |

**Table A15 List of most frequent words in the literature related to the neural correlates of AS extracted following the multivariate analysis pipeline.** The frequency of occurrence of each word was calculated across all papers for each individual brain region, and normalized between 0 and 1, with 1 corresponding to the most frequent word. The frequency scores of the 5 words with the highest values were added for all the brain regions related to AS. The neural correlates of AS were extracted following the multivariate analysis pipeline.

| Word | Cumulative frequency of occurrence |
|---|---|
| memory | 2.000000 |
| social | 1.679775 |
| behavior | 1.595449 |
| emotional | 1.332494 |
| learning | 1.281681 |
| fear | 1.221097 |
| schizophrenia | 1.099290 |
| auditory | 1.000000 |
| semantic | 1.000000 |
| anxiety | 1.000000 |
| reward | 1.000000 |
| speech | 0.980220 |
| priming | 0.849558 |
| pain | 0.840708 |
| reversal | 0.800448 |
| automatic | 0.752212 |
| age | 0.733766 |
| impairment | 0.727273 |
| development | 0.727273 |
| damage | 0.707792 |
| choice | 0.665919 |
| lesion | 0.609865 |
| stimulus | 0.530769 |
| expression | 0.504573 |
| crf | 0.472081 |
| subject | 0.470330 |
| synaptic | 0.419207 |
| stress | 0.297256 |